\documentclass[12pt,a4paper]{article}
\usepackage{latexsym}
\usepackage{amsmath}
\usepackage{amsfonts}
\usepackage{amsbsy}
\usepackage{amssymb}
\usepackage{graphicx}
\usepackage{mathrsfs}

\def \R{{\mathbb R}}
\def \C{{\mathbb C}}
\def \Z{\mathbb Z}

\def \D{{\mathscr D}}

\def \NN{\frac{\nu-\mu}{4\mu_0}}

\def \G{\Gamma}
\def \L{\Lambda}
\def \M{\mathcal{M}}

\def \a{\alpha}

\def \c{\chi}

\def \H{\mathcal{H}}
\def \d{\delta}

\def \l{\lambda}
\def \g{\gamma}
\def \r{\rho}
\def \s{\sigma}

\def \e{\varepsilon}

\def \Sym{\text{Sym}}
\def \z{\zeta}
\def \n{\nu}
\def \m{\mu}

\def \del{\partial}

\def \sgn{\rm sgn}

\def \nket{|\nu\rangle}
\def \nbra{\langle\nu|}
\def \mket{|\mu\rangle}
\def \mbra{\langle\mu|}

\numberwithin{equation}{section}

\begin{document}
\title{Approximating the physical inner product of Loop Quantum Cosmology\\[20pt]}

\author{\it \small Benjamin Bahr, MPI f\"ur Gravitationsphysik, Albert-Eintstein
Institut,\\ \it \small Am M\"uhlenberg 1, 14467 Golm, Germany\\[20pt]
\it \small Thomas Thiemann, MPI f\"ur Gravitationsphysik,
Albert-Eintstein Institut, \\ \it \small Am M\"uhlenberg 1, 14467
Golm, Germany;\\ \it \small Permineter Institute for Theoretical
Physics, \\ \it \small 31 Caroline St. N., Waterloo Ontario N2L
2Y5, Canada}

\maketitle

\abstract{\noindent In this article, we investigate the
possibility of approximating the physical inner product of
constrained quantum theories. In particular, we calculate the
physical inner product of a simple cosmological model in two ways:
Firstly, we compute it analytically via a trick, secondly, we use
the complexifier coherent states to approximate the physical inner
product defined by the master constraint of the system. We will
find that the approximation is able to recover the analytic
solution of the problem, which solidifies hopes that coherent
states will help to approximate solutions of more complicated
theories, like loop quantum gravity.\\[20pt]}

\section{Introduction}

\noindent Loop Quantum Gravity is an attempt to quantize General
Relativity to obtain a quantum theory of gravity (see
\cite{INTRO} and references therein). In order to do so, one first
has to rewrite GR as a Hamiltonian theory, which leads to a
constrained system. The question how to quantize a Hamiltonian
theory with constraints exists for a long time \cite{QGS, UNIQUE,
DIRAC}. The technical issues are, in fact, not too complicated.
The constraint functions $C_i$, which are phase space functions in
the classical theory, become operators $\hat C_i$ on an auxiliary,
or kinematical Hilbert space $\H_{kin}$. The classical issue of
solving the constraints is, on the quantum side, transformed into
the task of computing the common zero of the constraint operators
$\hat C_i\psi=0$. Unfortunately, since the point zero is
generically in the continuous parts of the spectra of the $\hat
C_i$, the space of solutions to the constraints will in general be
no subspace of the Hilbert space. Rather, the set of solutions
will contain distributions, which will not be normalizable in the
inner product on $\H_{kin}$. So, even when the space of solutions
of $\hat C_i\psi=0$ is known, it has no structure of a Hilbert
space yet. In fact, from various examples it is known that the
choice of a "physical" inner product is a quite nontrivial task,
and the resulting theory will crucially depend on this choice.
Generically, a lot of choices are possible which result in trivial
theories, or nontrivial theories that in no way resemble the
features of the classical theories one started with.

In Loop Quantum Gravity this issue is even more complicated, since
the algebra of constraints is tremendously difficult. Although it
is possible to define the Hamiltonian constraints \cite{QSD1,
QSD2} (which then still contain a lot of freedom), the
diffeomorphism constraints cannot be implemented as operators on
$\H_{kin}$ \cite{INTRO}. Furthermore, there are arguments that
even if such operators could be constructed, they could not be
self-adjoint, which makes spectral analysis quite difficult (see,
e.g. remarks in \cite{INTRO, PHOENIX}). Nevertheless, a
diffeomorphism invariant Hilbert space $\H_{diff}$ could be
defined, which contained vectors invariant under the exponentiated
diffeomorphism operators \cite{ALMMT}. But since the Hamiltonian
constraints do not even weakly commute with the diffeomorphism
constraints, there was no way of defining the Hamiltonian
constraints on $\H_{diff}$, so the search for the solutions to the
constraints could not continue.

In 2003, a solution to this set of problems was suggested
\cite{PHOENIX}. It is mainly motivated by the observation that a
classical constrained system with an arbitrary number of
constraints $C_i$ can be replaced by an equivalent system with
only one constraint, the so-called master constraint $M$. Solving
for all the $C_i$ simultaneously or solving for $M$ then leads to
identical solutions. In the corresponding quantum theories the
constraint operators $\hat C_i$ can be replaced by the master
constraint operator $\hat M$. Unfortunately, while on the
classical side these two sets of constraints lead to equivalent
classical theories, this equivalence is not that obvious in the
quantum theory. In generical situations, there are a lot more
solutions to $\hat M\psi=0$ than to $\hat C_i\psi=0$ for all $i$.
On the other hand, in generic models, even not all of the
solutions to the constraint, whenever one was able to compute
them, had non-vanishing physical norm, electromagnetism being the
first example. In many tested models, many of the formal solutions
of $\hat M=0$ also turned out to have zero norm in the physical
inner product \cite{TEST}. This reinforced the hope that the
master constraint could be used to define the physical inner
product and lead to sensible results: There was the possibility
that the additional solutions were all spurious, i.e. had all zero
physical norm.

Still, even with the master constraint, the task of computing the
physical inner product is, for the case of Loop Quantum Gravity,
technically very hard \cite{QSD8}. The operator $\hat M$ is quite
complicated and hence a spectral analysis is presumably not possible analytically. On the other hand,
this is nothing to be severely concerned about, since the
knowledge of the physical inner product would be equivalent to a
complete solution of the theory, i.e. a quantum theory which, in
its semiclassical sector, is believed to contain classical general
relativity. But even of classical GR one does not know all
solutions (in fact, one does not even know how many solutions one
doesn't know). So solving the complete theory is not something we
can expect at all. On the other hand, this is also not something
that is necessary in order to do physics. The situation is
similar to the case of condensed matter physics, where the
Hamiltonian operator for a crystal with roughly $10^{23}$
particles can be written down, but not one solution is known. And
this is not necessary in order to do physics, since no exact
solutions are needed to make predictions about conductivity,
magnetization or the shear module of a crystal.

So what one needs in order to extract physics from the equations
of LQG, are approximation schemes for the physical inner product
defined by $\hat M$. In this paper, we will investigate a simple
cosmological model with only one constraint $\hat C$ and try to
approximate the physical inner product defined by the
corresponding master constraint $\hat M$ with the help of
complexifier coherent states. These states have been introduced by
Hall \cite{HALL1, HALL2, HALL3} and investigated in \cite{GCS1, GCS2, GCS3}. They have properties that make
them useful tools for semiclassical analysis, in particular, they
will allow for an approximation for the physical inner product of
our toy model.

There are, with some tricks, other possibilities to calculate the
physical inner product in this model analytically \cite{PNV04},
without the need to employ the master constraint $\hat M$, and we
will see that the approximation provided by the coherent states in
fact works. Furthermore, the solutions to the master constraint
$\hat M$ that are no solutions to the constraint $\hat C$ turn out
not to be in the physical Hilbert space, as one hopes.

\section{Preliminaries}

\subsection{The master constraint}
\noindent Consider a phase space $\M$ and a set of first class
constraints $C_i$. This means that the Poisson brackets
 between two constraint functions $\{C_i,C_j\}$ vanish at the phase space
points where all the constraints $C_i$ vanish. Then the space of
all points $m\in\M$ with $C_i(m)=0$ for all $i$ is called
\emph{reduced phase space}. The Hamiltonian vector fields of the
constraints $X_{C_i}$ generate flows that leave this reduced phase
space invariant, and the orbit space is called the \emph{physical
phase space}. All phase space functions weakly commuting with the
constraints, called \emph{weak Dirac observables}, can then be
made into functions on the physical phase space, and will serve as
observables.

Instead of solving all the (possibly infinitely many) equations
$C_i(m)=0$, one can also define the so-called \emph{master
constraint} \cite{PHOENIX, QSD8}
\begin{align}\label{Gl:ClassicalMasterConstraint}
M\;:=\;\sum_iC_iK_{ij}C_i.
\end{align}

\noindent Here $K_{ij}$ is a symmetric, positive definite matrix
in the case of $i$ being a discrete set. Otherwise, $K_{ij}$ has
to be a positive definite operator kernel and the summation over
$i$ turns into an integration. It is straightforward to see that
\begin{align}\label{Gl:Kap2:DefinitionMasterConstraintClassical}
M(m)=0\;\qquad\Leftrightarrow\qquad\;C_i(m)=0\;\text{for all }i.
\end{align}

\noindent Furthermore, for any phase space function $f$ weakly
commuting with the constraints:
\begin{align}
\big\{\,\{M,\,f\},\,f\big\}\approx
0,\qquad\Leftrightarrow\qquad\;\{C_i,\,f\}\;\approx\;0\;\text{for
all }i.
\end{align}

\noindent So $M$ enables us to derive the complete set of
observables on the physical pase space. This means that the
physical phase space itself can be constructed from the knowledge
of $M$, so one does not lose any information if one goes over to
$M$ from the $C_i$. The final classical systems defined by both
are in fact equivalent. This is in fact independent of the actual
choice of $K_{ij}$, so there are a priori many possible master
constraints. One can choose the one that is most useful, makes the
sum (\ref{Gl:Kap2:DefinitionMasterConstraintClassical}) converge
and is the most convenient to compute. This freedom is quite
useful in the quantized theories \cite{TEST}.

\subsection{Quantizing constrained systems}

\noindent Consider a Hilbert space $\H_{kin}$ and a set of constraint
operators $\hat C_i$. Let furthermore $\D_{kin}\subset\H_{kin}$ be
a dense subspace which is invariant under the action of the $\hat
C_i$ and their adjoints $\hat C_i^{\dag}$. A \emph{solution to the
constraints} is a linear form
\begin{align}
l\,:\,\D_{kin}\,\to\,\C
\end{align}

\noindent with
\begin{align}
l\,\circ\,\hat C_i^{\dag}\;=\;0\qquad\mbox{for all }i
\end{align}

\noindent The space of all such linear forms $\D^*_{phys}$ is a
subspace of the algebraic dual of $\D_{kin}$:
\begin{align}
\D^*_{phys}\;\subset\;\D^*_{kin}.
\end{align}

\noindent In general, $\D^*_{phys}\cap\H_{kin}\,=\,\{0\}$, so the
solutions to the constraints will not be normalizable in the inner
product on $\H_{kin}$. Furthermore, not only can $\D^*_{phys}$ not
inherit the inner product from $\H_{kin}$, experience also shows
that not all elements in $\D^*_{phys}$ are acceptable as solutions
for generical models. Generally, $\D^*_{phys}$ is "too big" to
admit a positive definite inner product.

The task is now to find a Hilbert space $\H_{phys}$ that is a
subspace of $\D_{phys}^*$. In general, there will be a lot of
them, and the issue of finding the "right" one is a nontrivial
question. One necessary condition for this Hilbert space would for
instance be that it supports the quantizations of observables
weakly commuting with the constraints
$C_i$ as self-adjoint operators. \\

\noindent The idea is now to replace the set of constraint
operators $\hat C_i$ by the quantization of
(\ref{Gl:ClassicalMasterConstraint}), the  master constraint
operator
\begin{align}\label{Gl:Kap2:DefinitionMasterConstraintQuantum}
\hat M\;:=\;\sum_i\hat  C_i K_{ij} \hat C_i^{\dag}.
\end{align}

\noindent Here the freedom to choose $K_{ij}$ has in fact to be
exploited in order to arrive at a well-defined operator.

One can immediately see that $\D_{kin}$ is left invariant by $\hat
M$, and for any linear form $l\in\D_{kin}^*$:
\begin{align}\label{Gl:MoreSolutionsToMasterConstraint}
l\;\circ\;\hat C_i\;=\;0\;\quad\mbox{for all
}i\;\qquad\Rightarrow\qquad\;l\;\circ\;\hat M\;=\;0.
\end{align}

\noindent If $0$ is only contained in the discrete (point) spectrum
of all the $\hat C_i$, then (\ref{Gl:MoreSolutionsToMasterConstraint})
is an equivalence relation. But if $0$ is in the continuous part of
the spectrum, then the converse of  then (\ref{Gl:MoreSolutionsToMasterConstraint})
need not be true. So the space of
solutions of $\hat M$ is potentially larger than the space of
solutions to the constraints $\hat C_i$. On the other hand, $\hat
M$ provides a natural way to define the physical Hilbert space. In
all models checked so far, this Hilbert space turned out to
contain no solutions of $\hat M$ that were no solutions to the
$\hat C_i$. Although there are no hard proofs available at the
moment, there are high hopes that this is the case generally.


\subsection{The physical Hilbert space and rigging maps}\label{Ch:PHYS}

\noindent Since $\hat M$ is formally symmetric and non-negative by
construction, one can always extend it to a self-adjoint operator.
Of course, this has to be checked from case to case, and we assume
it to be true here. Then $\hat M$ is also positive and hence its
spectrum is a subset of $[0,\infty)$. We furthermore assume that
$0$ is in fact in the spectrum of $\hat M$. Then, for each $\l\in
spec(\hat M)$ there is a Hilbert space $\H_{\l}$, equipped with
inner product $\langle\,\cdot\,|\,\cdot\,\rangle_{\l}$.
Furthermore there is a measure $\m$ on $spec(\hat M)$ such that
each vector of $\H_{kin}$ can be decomposed into
\begin{align}
\psi\;=\;\int_{spec(\hat M)}d\m(\l)\;\psi(\l)
\end{align}

\noindent and the inner product in $\H_{kin}$ via
\begin{align}
\langle \psi|\phi\rangle\;=\;\int_{spec(\hat M)}d\m(\l)\,
\big\langle\,\psi(\l)\,\big|\,\phi(\l)\,\big\rangle_{\l}.
\end{align}

\noindent These facts can be abbreviated by
\begin{align}\label{Gl:Kap2:DID}
\H_{kin}\;=\;\int_{spec(\hat M)}^{\oplus}d\m(\l)\;\H_{\l}
\end{align}

\noindent which is called the \emph{direct integral decomposition}
of $\H_{kin}$. We neglect all discussions about the mathematical
subtleties of this construction and just state the fact that the
Hilbert space $\H_{\l}$ for $\l=0$ shall serve as physical Hilbert
space.
\begin{align}\label{Gl:Kap2:DID2}
\H_{phys}\;:=\;\H_{0},\qquad\langle\,\cdot\,|\,\cdot\,\rangle_{phys}\;:=\;\langle\,\cdot\,|\,\cdot\,\rangle_{0}.
\end{align}

\noindent This procedure can in principle be applied to every
system with constraints. In particular, in the case of $0$ being
in the discrete part of the spectrum only, this is just the
definition of zero eigenspace of $\hat M$. In the case of a mixed
spectrum of $\hat M$, one has to decompose $\H_{kin}$ into a
"discrete" and "continuous" part and perform this analysis for
each part separately:
\begin{align}
\H_{kin}\;=\;\H_{kin}^{pp}\,\oplus\,\H_{kin}^{ac}\;=\;\int_{spec_{pp}(\hat
M)}^{\oplus}d\m_{pp}(\l)\;\H^{pp}_{\l}\;\oplus\;\int_{spec_{ac}(\hat
M)}^{\oplus}d\m_{ac}(\l)\;\H^{ac}_{\l}.
\end{align}

\noindent Here "pp" and "ac" denote the pure-point- and the
absolutely continuous part of the spectrum of $\hat M$. In the
end, the physical solutions from each separate part constitute the
whole physical Hilbert space, in particular
\begin{align}\label{Gl:Blablabla}
\H_{phys}\,=\,\H_{0}^{pp}\,\oplus\,\H_{0}^{ac}.
\end{align}

\noindent Note that $\H^{pp}_0\subset \H_{kin}$ consists of the
zero eigenvectors of $\hat M$. For both cases "pp" and "ac" the
Hilbert space $\H_0$ is a subspace of $\D_{phys}^*$ in the
following sense: In the direct integral decomposition
$\psi\in\H_{kin}$ corresponds to a map $\l\mapsto\psi(\l)$ with
$\psi(\l)\in\H_{\l}$. Denote the evaluation of this map at $\l=0$
as $\psi(0)$. Then define for each $\psi\in\H_{kin}$ the following
linear functional:
\begin{align}\label{Gl:Kap2:RiggingMapAbstract}
\eta[\psi]\;:\;\phi\;\longmapsto\;\big\langle\psi(0)\big|\phi(0)\big\rangle_0.
\end{align}

\noindent So $\eta$ provides a "projection" of $\H_{kin}$ in
$\H_0$ (which is a proper projection for the pure-point case). On
the other hand, since the decomposition of $\hat M\phi$ is given
by the map $\l\mapsto \l\phi(\l)$, it is immediately clear that
$\eta$ as defined in (\ref{Gl:Kap2:RiggingMapAbstract}) maps
$\psi$ into a solution of the constraint $\hat M$, i.e.
$\eta[\psi]\circ\hat M=0$. In the literature one also finds for
$\eta$ the name
\emph{rigging map}.\\

\noindent Although conceptually quite clear, the decomposition
(\ref{Gl:Kap2:DID}) is in general nontrivial, as the spectral
analysis for an arbitrary self-adjoint operator can be highly
complicated. This is also the fact for the master constraint
operator of LQG. Although the operator has been constructed
explicitly \cite{QSD8}, its complete spectral analysis is currently out of reach.\\

\subsection{The operator $\d(\hat M)$}

\noindent We now would like to find a way of approximating the
solutions of $\hat M$, i.e. the linear functionals
(\ref{Gl:Kap2:RiggingMapAbstract}) for all $\psi\in\H_{kin}$. For
this we observe
\begin{align}\nonumber
\eta[\psi]\phi\;&=\;\;\big\langle\psi(0)\big|\phi(0)\big\rangle_0\\[5pt]\label{Gl:Kap2:RewritingRiggingMapWithDeltaOfM}
&=\;\int_{spec(\hat
M)}d\m(\l)\;\big\langle\psi(\l)\big|\d(\l)\big|\phi(\l)\big\rangle_{\l}\\[5pt]\nonumber
&=\;\big\langle\psi\big|\d(\hat M)\big|\phi\big\rangle,
\end{align}

\noindent where $\d(\hat M)$ depends on $d\m(\l)$. The "operator" $\d(\hat M)$ is, of course, no operator
but a quadratic form which will not exist for all
$\psi,\,\phi\in\H_{kin}$, but only for those vectors, whose
decomposition into $\l\mapsto\psi(\l)$ is regular at $\l=0$.\\

\noindent Equation (\ref{Gl:Kap2:RewritingRiggingMapWithDeltaOfM})
now enables us to express the linear form $\eta[\psi]$ without
directly computing the direct integral decomposition. All we need
is the construction of a $\d$-function for the measure $\m$ on the
spectrum of $\hat M$. For this, the knowledge of the measure $\m$
is, of course, sufficient, but not necessary. All one needs to
know is the rate at which the measure "diverges" or "goes to zero"
at $\l=0$.

If $0$ is in the discrete part of the spectrum, the measure $\m$
is a Dirac-measure there, and the delta-function is just a usual
Kronecker-delta. In this case the integral (\ref{Gl:Kap2:DID}) is
just a sum over $\l=0$ and hence $\d(\hat M)$ is a proper
projection operator which maps $\H_{kin}$ onto its subspace
$\H^{pp}_0$.

If $0$ is in the continuous part of the spectrum, the measure $\m$
can behave very singularly at $\l=0$. This is supported by the
following observation: Consider the didirect integral decomposition
 of a self-adjoint operator $\hat A$, whose spectrum is
the real line, and whose spectral measure $\m_{\hat A}$ is the
Lebesgue measure $d\l$. Then the operator $\hat A^2$ has
$[0,\infty)$ as spectrum, and its measure $\m_{\hat A^2}$ is then
$\frac{1}{2\sqrt{\l}}d\l$, which diverges as $\l$ tends to zero.
So by taking powers of operators, one changes the behavior
of the spectral measure at $\l=0$.\\

\noindent So, an educated guess for a delta-sequence are the
functions
 \begin{align}
\l\;\longmapsto\;t^{-\a}\,e^{-\frac{\l^2}{t}}.
 \end{align}

 \noindent If the parameter $\a\in\R$ is chosen correctly, so as to
match the degree of divergence of the measure $\m$ at $\l=0$, then
this diverges to a multiple of the $\d$-distribution at $\l=0$, as
$t$ tends to $0$. Determining $\a$ can be nontrivial. On the other
hand, there is a different way of calculating the physical inner
product without the need to know $\a$ explicitly: Choose a vector
$\psi\in\H_{kin}$ such that in its direct integral decomposition
one has
\begin{align}\label{Gl:Kap2:NormConditionForReferenceVector}
0\;<\;\big\|\psi(0) \big\|_0\;<\;\infty.
\end{align}

\noindent Then one has
\begin{align}\nonumber
\lim_{t\to
0}\;\frac{e^{-\frac{\hat M^2}{t}}}{\langle\psi|e^{-\frac{\hat
M^2}{t}}|\psi\rangle}\;& =\;\lim_{t\to
0}\;\frac{t^{-\a}\;e^{-\frac{\hat M^2}{t}}}{t^{-\a}\;\langle\psi|e^{-\frac{\hat
M^2}{t}}|\psi\rangle}\;=\;\lim_{t\to
0}\;\frac{t^{-\a}\;e^{-\frac{\hat M^2}{t}}}{t^{-\a}\;\;\int
d\mu(\l)\langle\psi(\l)|e^{-\frac{
\l^2}{t}}|\psi(\l)\rangle_{\l}}\\[5pt]\label{Gl:RiggingMapEquivalence}
&=\;\frac{\lim_{t\to 0}\;t^{-\a}\;e^{-\frac{\hat M^2}{t}}}{\lim_{s\to
0}\;s^{-\a}\;\;\int d\mu(\l)\langle\psi(\l)|e^{-\frac{
\l^2}{s}}|\psi(\l)\rangle_{\l}}\\[5pt]\nonumber
&=\;\frac{1}{\big\|\psi(0)\big\|_0}\;\d(\hat M).
\end{align}

\noindent Here, the equalities have to be understood in the sense
of quadratic forms. This shows that, by employing a "reference
vector" $\psi$ one can save oneself from the need to explicitly
calculating $\a$. The resulting physical inner product is then
rescaled such that $\eta[\psi]$ has unit norm in the physical
inner product. One now has shifted the problem of finding $\a$ to
finding a reference vector that satisfies
(\ref{Gl:Kap2:NormConditionForReferenceVector}). But since the set
of vectors satisfying
(\ref{Gl:Kap2:NormConditionForReferenceVector}) is dense in
$\H_{kin}$ for $\l=0$ being in the the continuous spectrum of
$\hat M$, one would think that finding such a vector is not too
complicated.


\section{The model}\label{Ch:TheModel}

\noindent As a test of full LQG one can focus attention to symmetry
truncations of LQG, in the hope to get technical insight in soluable
situations which are still valid in the full theory. A particular
class of truncations modelling cosmological situations, called Loop
Quantum Cosmology (LQC), has received a
lot of interest recently \cite{ABL03}. We will apply the
approximation scheme we have in mind to the easiest LQC-model
possible: that of a spatially homogenous and isotropic universe
without any matter fields. For the case of a nonzero cosmological
constant, the physical Hilbert space has been constructed in
\cite{PNV04}. The case for vanishing cosmological constant is
different, since then the zero is not in the discrete part of the
spectrum any longer. It is this case that we will investigate in
this article.

The metric the model describes is
\begin{align}
ds^2\;=\;-dt^2\,+\,a(t)^2\,(dx^2+dy^2+dz^2).
\end{align}

\noindent The canonical variables that resemble the Ashtekar
variables in full LQC are
\begin{align}
p\;=\;a^2,\qquad c\;=\frac{\dot a}{2a},
\end{align}

\noindent which have
\begin{align}
\big\{\,c,\,p\,\big\}\;=\;1.
\end{align}

\noindent In the case of Loop Quantum Cosmology, $p$ is allowed to
actually take values in $\R$, so as to allow for a change in
orientation. Note that this normalization differs from the
literature by a factor of $8\pi\g G_N/3$ \cite{ABL03}.

The constraint of the system is given by
\begin{align}\label{Gl:TheConstraint}
C\;=\;-24\,\sgn(p)\cdot\sqrt{|p|}\,c^2\;+\;3\L\sqrt{|p|}^{\frac{3}{2}}.
\end{align}

\noindent  The kinematical Hilbert space $\H_{kin}$ of LQC is the space of almost periodic functions in $c$, which
has the functions
\begin{align}\label{Gl:BasisForHilbertSpace}
|\n\rangle\;:=\;c\,\longmapsto\,e^{i\frac{\n}{2} c},\qquad \n\in\R
\end{align}

\noindent as a basis. So $\H_{kin}$ is not separable, since it has
an uncountable basis, a feature that it shares with the Hilbert
space of full Loop Quantum Gravity. The quantization of $p$ acts
on $\H_{kin}$ via differentiation by $c$:
\begin{align}\label{Gl:EigenvectorOfP}
\hat p \,|\n\rangle\;=\;\frac{\n}{2}\,|\n\rangle,
\end{align}

\noindent whereas $c$ is not definable as an operator. Rather, one
has to employ $\exp(i\l\hat c)$ for all $\l\in\R$ as
multiplication with the corresponding function of $c$, via:
\begin{align}
e^{i\l\hat c}\,|\n\rangle\;=\;|\l+2\n\rangle,
\end{align}

\noindent which shows that the failing weak continuity of
$\l\mapsto\exp(i\l\hat c)$ is the reason why one cannot define
$\hat c$ without the exponential. This amounts to a problem when
attempting to define the quantization of (\ref{Gl:TheConstraint}),
since it explicitly contains $c$. This problem occurs in a similar
fashion in the full theory, and it is resolved in LQC by introducing a
parameter $\m_0$ and approximating $c^2$ by
$\m_0^{-2}\sin^2(\m_0c)$. \footnote{In full LQG the parameter $\m_0$
is not necessary, because spatial diffeomorphism invariance "swallows
it up" there. In LQC the spatial diffeomorphism group has been gauge-fixed
so that one has to deal with $\m_0$ as an artifact of the model}

If $c\ll 1/\m_0$, the two functions are
quite similar, thus, if $\m_0$ is sufficiently small, the function
will be a good substitute for a large part of phase space. The
latter function, however, can be quantized on $\H_{kin}$, which
results in the following constraint operator:
\begin{align}
\hat C\;=\;-24\,\left(\frac{\sin\mu_0 \hat
c}{\mu_0}\right)^2\sgn\,\hat
p\;\widehat{\sqrt{|p|}}\,\;+\;3\L\widehat{\sqrt{|p|}}^{\frac{3}{2}},
\end{align}

\noindent where
\begin{align}\label{Gl:ActionOfVolumeOperator}
\sgn\,\hat
p\;\widehat{\sqrt{|p|}}\;|\n\rangle\;=\;\frac{|\n+\m_0|^{\frac{3}{2}}\,-\,|\n-\m_0|^{\frac{3}{2}}}{3\m_0}\;|\n\rangle
\end{align}

\noindent is the quantization of $\sgn\,p\;\sqrt{|p|}$, that
results from the attempt to keep the quantization scheme as close
as possible to the ones employed in Loop Quantum Gravity.

The introduction of the scale $\m_0$ induces a splitting:
\begin{align}
\H_{kin}\;=\;\bigoplus_{\d\in[0,4\m_0)}\H_{\d}
\end{align}

\noindent where each $\H_{\d}\,=\,{\rm
span}\,\{|\n\rangle,\,\n\in\d+4\m_0\Z\}$ is separable and left
invariant by the constraint. Thus, one can calculate the rigging
map for each sector $\d\in[0,4\m_0)$ separately.\\


\subsection{Solutions to the Constraint}\label{Ch:KapitelEins}

\noindent We attempt to calculate the physical inner product for
the above cosmological model with $\L=0$.  For this, we first use
a trick that has been employed in \cite{PNV04} to compute the
physical inner product for the case with $\L\neq 0$ of
\emph{Riemannian} cosmology. For this case the 0 is in the pure
point spectrum of the constraint, and hence the eigenvalue problem
could be solved explicitly. It was reasoned that this stays true
for the case of Lorentzian cosmology with nonzero cosmological
constant. If $\L$ is set to zero, then Riemannian and Lorentzian
cosmology become indistinguishable, and then $0$ is not any longer in
the pure point spectrum of $\hat C$. Still, the trick used in
\cite{PNV04} will make it possible to compute the physical Hilbert
space analytically.

Since the trick is in principle not applicable in full Loop
Quantum Gravity, we will also try to calculate the physical inner
product without using it. Furthermore, since without the trick the
constraint operator is no longer symmetric, we work with the
master constraint operator $\hat M=\hat C^{\dag}\hat C$ instead.
In this setting, the master constraint operator will be too
complicated to derive the physical inner product analytically, but
the coherent states will enable us to approximate it well enough. This is precisely what one expects also in full LQG
In particular, we will find the approximate solutions being close
to the solutions previously obtained analytically, which will show
that the coherent states are in principle useful to approximate
physical inner products. It will also demonstrate that, although
the original constraint operator $\hat C$ has been replaced by
$\hat M$, which has potentially many more solutions, the physical
Hilbert space computed from $\hat M$ consists of solutions of
$\hat C$ only.

\subsection{Riemannian cosmology with and without cosmological
constant}\label{Ch:RiemannianCosmology}

\noindent We shortly repeat the results of \cite{PNV04}: There,
the Hilbert space was the same that the one used in the Lorentzian
case we are looking at, but the constraint looked differently. In
particular, it was given classically by
\begin{align}\label{Gl:Riemannian1}
C_{\rm Riemann}\;=\;24\,\sgn
p\cdot\sqrt{|p|}\,c^2\;+\;3\L\sqrt{|p|}^{\frac{3}{2}}.
\end{align}

\noindent The trick is to modify this constraint by multiplication
with $|p|^{-\frac{1}{2}}$, to arrive at the (up to the points
$p=0$) classically equivalent function

\begin{align}\label{Gl:Riemannian2}
C_{\rm Riemann}\;=\;24\,c^2\;+\;3\L p.
\end{align}

\noindent By introducing a length parameter $\m_0$, as shown in
section (\ref{Ch:TheModel}), this function can be quantized on
$\H_{kin}$ as

\begin{align}
\hat C_{\rm Riemann}=24\,\left(\frac{\sin\mu_0 \hat
c}{\mu_0}\right)^2\;+\;3\L\hat p,
\end{align}

\noindent from which the physical inner product was computed in
\cite{PNV04}. The only solution to  (\ref{Gl:Riemannian2}), which
has nonzero physical norm, was calculated to be
\begin{align}\label{Gl:PhysicalSolutionRiemann}
l\;=\;\sum_{\l\in-\frac{6}{\L\mu_0^2}+4\mu_0\Z}\;J_{\frac{\l}{4\mu_0}-\frac{3}{2\L\mu_0^3}}\left(\frac{3}{2\L\mu_0^3}
\right)\;\langle\l|,
\end{align}

\noindent where $J_n(x)$ is the $n-$th Bessel function of first
kind. In particular, although there exist solutions to the
constraint in every sector $\d\in[0,4\mu_0)$, only the solution in
the sector $\d=4\mu_0[1-\mod(\frac{3}{2\L\mu_0^3},1)]$ has nonzero
physical norm. Here,$\mod(x,\,y)$ denotes the remainder of $x/y$. So by the constraint, one sector is allowed, and
the resulting physical Hilbert space is one-dimensional, as
assumed for a theory with zero degrees of freedom.

If one lets $\L\to 0$,  neither does the solution
(\ref{Gl:PhysicalSolutionRiemann}) converge, nor does the sector $\d$
in which it is defined. So, in order to investigate the case
$\L=0$, we have to calculate the physical inner product
explicitly, instead of taking the limit $\L\to 0$ of the already
known solutions. Classically, this amounts to
\begin{align}\label{Gl:SimpleConstraintClassical}
C\;=\;24\,c^2.
\end{align}

\noindent Since the quantized constraint
\begin{align}\label{Gl:SimpleConstraint}
\hat C=24\,\frac{\sin^2\mu_0 \hat c}{\mu_0^2}
\end{align}

\noindent is self-adjoint, we do not employ the master constraint,
but calculate the rigging map (\ref{Gl:RiggingMapEquivalence})
directly with $\hat C$. We have to compute
(\ref{Gl:RiggingMapEquivalence})

\begin{align*}
\eta[\nket]\mket&\;=\;\lim_{t\to 0}\frac{\mbra\,\,e^{- \frac{\hat
C^2}{t}}\,\,\nket}{\langle \n_0\,|\,e^{- \frac{\hat
C^2}{t}}\,|\,\n_0\rangle},
\end{align*}

\noindent  where $|\n_0\rangle$ is a reference vector for correct
normalization.

 Since $\hat C$ is bounded, we have
\begin{align*}
\nbra\,e^{- \frac{\hat
C^2}{t}}\,\mket&\;=\;\sum_{n=0}^{\infty}\frac{(-1)^n}{n!}\;\frac{1}{t^{n}}\,\mbra
\hat C^{2n}
\nket\\[5pt]\nonumber
&\;=\;\sum_{n=0}^{\infty}\frac{(-1)^n}{n!}\;\left(\frac{24}{\mu_0^2\sqrt
t}\right)^{2n}\,\mbra \sin^{4n}(\mu_0\hat c) \nket.
\end{align*}

\noindent With definition of $e^{i\m_0\hat c}$ one can see that
\begin{align}
\mbra \sin^{4n}(\mu_0\hat c) \nket\;=\;\left\{\begin{array}{ll}
\frac{1}{16^n}\,(-1)^{\frac{\nu-\mu}{4\mu_0}}\,\left(\begin{array}{c}4n\\2n+\frac{\nu-\mu}{4\mu_0}\end{array}\right)&
\qquad\text{for }\frac{\nu-\mu}{4\mu_0}\,\in\,\Z\\ 0&\qquad
\text{else} \end{array}\right.
\end{align}

\noindent So in what follows, assume $\frac{\nu-\mu}{4\mu_0}$ to
be integer. Then we have
\begin{align}\nonumber
\nbra\,e^{- \frac{\hat
C^2}{t}}\,\mket&\;=\;(-1)^{\frac{\nu-\mu}{4\mu_0}}\,\sum_{n=0}^{\infty}\frac{(-1)^n}{n!}\;\left(\frac{6}{\mu_0^2\sqrt
t}
\right)^{2n}\,\left(\begin{array}{c}4n\\2n+\frac{\nu-\mu}{4\mu_0}\end{array}\right)\\[5pt]\nonumber
&\;=\;(-1)^{\frac{\nu-\mu}{4\mu_0}}\,\sum_{n=0}^{\infty}\frac{(-1)^n}{n!}\;\left(\frac{6}{\mu_0^2\sqrt
t}\right)^{2n}\,
\frac{\G(4n+1)}{\G(2n+1+\frac{\nu-\mu}{4\mu_0})\,\G(2n+1-\frac{\nu-\mu}{4\mu_0})}\\[5pt]\label{Gl:FormelFuerMellinBarnes}
&\;=\;(-1)^N\,\sum_{n=0}^{\infty}\frac{(-1)^n}{n!}\;a^{2n}\,\frac{\G(4n+1)}{\G(2n+1+N)\,\G(2n+1-N)}
\end{align}

\noindent with the abbreviations
\begin{align*}
N\,=\,\frac{\nu-\mu}{4\mu_0},\qquad a\,=\,\frac{6}{\mu_0^2\sqrt
t}.
\end{align*}

\noindent Now we use the identity
\begin{align}\label{Gl:ChristiansFormelohneSinus}
\sum_{n=0}^{\infty}\frac{(-1)^n}{n!}\,f(n)\;=\;\frac{1}{2\pi
i}\int_{-i\infty}^{i\infty}ds\;\G(-s)\,f(s),
\end{align}

\noindent (see e.g. \cite{ABR}, p.559), for $f$ being as in (\ref{Gl:FormelFuerMellinBarnes}).  A coordinate
transformation $s=-\frac{r+1}{4}$ yields

\begin{align*}
\sum_{n=0}^{\infty}\frac{(-1)^n}{n!}&\,a^{2n}\,\frac{\G(4n+1)}{\G(2n+1+N)\,\G(2n+1-N)}\\[5pt]
&\;=\;\frac{1}{2\pi i}\int_{-i\infty}^{i\infty}ds\;\G(-s)\,a^{2s}\,\frac{\G(4s+1)}{\G(2s+1+N)\,\G(2s+1-N)}\\[5pt]
&\;=\;-\frac{1}{8\pi
i}\int_{i\infty+1}^{-i\infty+1}dr\;\G(-r)\,a^{-\frac{r+1}{2}}
\,\frac{\G\left(\frac{r+1}{4}\right)}{\G\left(\frac{1}{2}-\frac{r}{4}+N\right)\,\G\left(\frac{1}{2}-\frac{r}{4}-N\right)}.
\end{align*}

\noindent Using again (\ref{Gl:ChristiansFormelohneSinus}), we get

\begin{align*}
\sum_{n=0}^{\infty}\frac{(-1)^n}{n!}&\,a^{2n}\,\frac{\G(4n+1)}{\G(2n+1+N)\,\G(2n+1-N)}\\[5pt]
&\;=\;\frac{1}{4}\,a^{-\frac{1}{2}}\,\sum_{n=0}^{\infty}\frac{(-1)^n}{n!}\;a^{-\frac{n}{2}}\;
\frac{\G\left(\frac{n+1}{4}\right)}{\G\left(\frac{1}{2}-\frac{n}{4}+N\right)\,\G\left(\frac{1}{2}
-\frac{n}{4}-N\right)}.
\end{align*}

\noindent So, finally:
\begin{align*}
\nbra\,e^{- \frac{\hat
C^2}{t}}\,\mket&\;=\;\frac{(-1)^{\frac{\nu-\mu}{4\mu_0}}}{4}\;\;\left(\frac{\mu_0^2\sqrt
t}{6}\right)^{\frac{1}{2}}\,
\sum_{n=0}^{\infty}\frac{(-1)^n}{n!}\;\left(\frac{\mu_0^2\sqrt
t}{6}\right)^{\frac{n}{2}}\; \,\frac{\G\left(\frac{n+1}{4}\right)}
{\G\left(\frac{1}{2}-\frac{n}{4}+\frac{\nu-\mu}{4\mu_0}\right)\,\G\left(\frac{1}{2}-\frac{n}{4}-\frac{\nu-\mu}
{4\mu_0}\right)}.
\end{align*}

\noindent With this, (\ref{Gl:RiggingMapEquivalence}) and
\begin{align*}
\frac{1}{\G\left(\frac{1}{2}+N\right)\,\G\left(\frac{1}{2}-N\right)}\;=\;\frac{1}{\pi}\,(-1)^N
\end{align*}

\noindent we get
\begin{align}
\eta[\nket]\,\mket\;=\;\lim_{t\to 0}\;\frac{\nbra\,e^{- \frac{\hat
C^2}{t}}\,\mket} {\langle \n_0|\,e^{- \frac{\hat
C^2}{t}}\,|\n_0\rangle}\;=\;1
\end{align}

\noindent if $\frac{\n-\m}{4\mu_0}$ is an integer, $0$ else for
any $|\nu_0\rangle\in\H_{kin}$. Since the result does not depend
in the choice of $|\nu_0\rangle$, we might take any one.

From this we see the following: The physical Hilbert space is not
separable, it consists of the linear forms
\begin{align*}
l_{\d}\;:=\;\sum_{\l\in\d+4\mu_0\Z}\,\langle\l|,\qquad\d\in[0,4\m_0)
\end{align*}

\noindent with the physical inner product
\begin{align*}
\langle\,l_{\d}\,|\,l_{\e}\,\rangle_{phys}\;=\;\d_{\d\e}.
\end{align*}

\noindent So it decomposes into uncountably many one-dimensional,
orthogonal subspaces:
\begin{align*}
\H_{phys}\;=\;\bigoplus_{\d\in[0,4\m_0)}\C.
\end{align*}

\noindent This shows an important fact: Although the classical
system under consideration has zero degrees of freedom, the
corresponding physical Hilbert space $\H_{phys}$ of the quantum
theory has uncountably many dimensions. Also, the theory is not
superselected, since the operators
\begin{align}
\hat O_{\l} := \sin(\l \hat c),\qquad \l>0
\end{align}

\noindent are self-adjoint (with respect to the physical inner
product) and commute with the constraint, hence turn into Dirac
observables on the physical Hilbert space. These observables act
irreducibly in $\H_{phys}$, hence the physical Hilbert space is
not superselected.\\

The fact that the quantum theory behaves so significantly
different from the classical one can be traced back to the
quantization procedure. Since the constraint, the phase space
function $c^2$, has no quantization on $\H_{kin}$, one has to come
up with a different operator and chooses $\m_0^{-2}\sin^2 \m_0\hat
c$ for some fixed $\m_0$. This $\m_0$ introduces a scale into the
theory that has not been there before. Had one chosen the
uncountable set of constraints $\l^{-2}\sin^2 \l\hat c$ for
$\l>0$, which is classically equivalent to $c=0$, then the only solution for this system would have been the
linear form
\begin{align}
l\;=\;\sum_{\n\in\R}\langle\n|.
\end{align}

\noindent In this case the space of solutions to the constraints
had been one-dimensional right from the beginning, but the
definition of the physical inner product would have been harder
(though unnecessary in this simple case).

Note that the unphysically large size of $\H_{phys}$ should be
considered as an artifact of the model, resulting from the
introduction of the scale $\m_0$ to a formerly scale-invariant theory, i.e. going over from  (\ref{Gl:SimpleConstraintClassical})
to (\ref{Gl:SimpleConstraint}). This feature does not occur,
if the model already containes scales, i.e. determined by massive matter fields or
the cosmological constant $\L$ (as, for instance in \cite{PNV04}, where
the physical Hilbert space is onedimensional, as expected). Still, the results, of
course, depend on the relation of the scales from the theory and the scale
from $\m_0$, as one can see i.e. in (\ref{Gl:PhysicalSolutionRiemann}).


\subsection{The Lorentzian case}
\noindent The computation of the physical inner product in the
case of vanishing cosmological constant has been possible
explicitly, because the constraint (\ref{Gl:Riemannian1}) had been
changed into the classically equivalent (\ref{Gl:Riemannian2}).
Unfortunately, this trick is not available for the full theory,
for some important reason: To become an honest operator in full
Loop Quantum Gravity, the density weight of a phase space function
has to be $+1$! But multiplying with a nonzero power of the triads
$E_i^a$ (which correspond to $p$ in the cosmological case) changes
the density weight, hence destroys the quantization scheme. The
honest constraint is therefore
\begin{align}\label{Gl:NotSoSimpleConstraint}
\hat C \;=\;\left(\frac{\sin \m_0\hat
c}{\m_0}\right)^2\;\sgn\,\hat p\,\widehat{\sqrt{|p|}},
\end{align}

\noindent We would like to consider this constraint instead of
(\ref{Gl:SimpleConstraint}). In this form, the physical inner
product can no longer be calculated explicitly, but we will show
how the complexifier coherent states will be able to provide an
approximation scheme for it.\\

We start by giving all solutions of the constraint equation
\begin{align}\label{Gl:ConstraintEquation}
l\circ\hat C^{\dag}=0.
\end{align}

\noindent In particular, the solutions are, for $\d\in(0,4\mu_0)$:
\begin{align}\label{Gl:SolutionsDelta}
l_{(a,b)}\;=\;\sum_{\l\in\d+4\mu_0\Z}\frac{a+b\l}{|\l+\mu_0|^{\frac{3}{2}}\,-\,|\l-\mu_0|^{\frac{3}{2}}}\,
\langle\l|,\qquad
a,b\in\C,
\end{align}
\noindent whereas the solutions in the sector $\d=0$ are the
following:
\begin{align}\label{Gl:SolutionsDeltaZero}
l_{(a,b)}\;=\;\sum_{\l\in\d+4\mu_0\Z}\frac{a\l}{|\l+\mu_0|^{\frac{3}{2}}\,-\,|\l-\mu_0|^{\frac{3}{2}}}\,\langle\l|\;
+\;b\langle
0|,\qquad a,b\in\C.
\end{align}

\noindent Notice that the space of solutions is two-dimensional in
each case, consisting of one bounded and one unbounded solution.
This means that the set of coefficients $l_{(a,b)}|\l\rangle$ in (\ref{Gl:SolutionsDelta})
and (\ref{Gl:SolutionsDeltaZero}) are either bounded or unbounded.
The bounded solutions are normalizable in the kinematical inner
product for the case $\d=0$ only!

 The Master constraint is quite simple, since there is
only one constraint:
\begin{align}\label{Gl:MasterConstraint}
\hat M\;=\;\hat C^{\dag}\hat C,
\end{align}

\noindent i.e.

\begin{align}\label{Gl:MasterConstraintPowersExpectationValues}
\hat M\;=\;\widehat{\sqrt{|p|}}\,\sgn\,\hat p \;\left(\frac{\sin
\m_0\hat c}{\m_0}\right)^4\;\sgn\,\hat p\,\widehat{\sqrt{|p|}}
\end{align}

\noindent From this form it is immediately clear that the
solutions (\ref{Gl:SolutionsDelta}) and
(\ref{Gl:SolutionsDeltaZero}) satisfy
\begin{align}\label{Gl:MasterConstraintEquation}
l_{(a,b)}\circ\hat M=0.
\end{align}
\noindent Still, there are more solutions to
(\ref{Gl:MasterConstraintEquation}). These are, in particular, for $\d\in(0,\,4\m_0)$:

\begin{align}\label{Gl:AdditionalSolutions}
\tilde
l_{(c,d)}\;=\;\sum_{\l\in\d+4\mu_0\Z}\frac{c\l^2\,+\,d\l^3}{|\l+\mu_0|^{\frac{3}{2}}\,-\,|\l-\mu_0|^{\frac{3}{2}}}\,\langle\l|,
\qquad c,\,d\in\C.
\end{align}

\noindent The linear forms (\ref{Gl:AdditionalSolutions}) solve (\ref{Gl:MasterConstraintEquation}), but not
(\ref{Gl:ConstraintEquation}). So, as one would have guessed from
the outset, there are more solutions to the Master Constraint than
to the original constraint, and it is common belief that the
additional solutions will turn out to be spurious, that is, have
zero norm in the physical inner product yet to compute.

The form of the master constraint
(\ref{Gl:MasterConstraintPowersExpectationValues}) is too complicated to give a closed expression for
$\nbra\,\hat M^{2n}\,\mket$, which is needed in order to compute
the rigging map (\ref{Gl:RiggingMapEquivalence}). It is at this
point, where the coherent states come into play, since they
  allow for an approximation of
 (\ref{Gl:MasterConstraintPowersExpectationValues}), as will be
 shown in the following.


\subsection{Coherent states}

\noindent The complexifier coherent states have been defined in
\cite{HALL1, HALL2, HALL3} and their properties have been
investigated in \cite{GCS1, GCS2, GCS3, CCS}, where their
particular use for approximations has been pointed out. We will
use the simplest of these states, in particular the ones for the
gauge group $U(1)$. They are defined on each $\H_{\d}\;=\;{\rm
span}\;\{|\l\rangle\,,\;\l\in \d+4\mu_0\Z\}$, with
$\d\in[0,4\mu_0)$, by
\begin{align}\label{Gl:CoherentStates}
|u\rangle\;=\;\sum_{\l\in\d+4\m_0\Z}e^{-\l^2\frac{s^2}{2}}\,e^{i\l
u}\,|\l\rangle.
\end{align}

\noindent The complex number $u=c-ip$ that labels the coherent
states resemble the phase-space coordinates. With our conventions
here, the states are peaked in phase-space around the points
$(-2c, p/s^2)$ with width $s$. Hence, $s$ plays the role of a "classicality parameter".

The coherent states have a number of properties that make them
useful for approximations. In particular, they provide a
resolution of unity via

\begin{align}\label{Gl:RichtigeZerlegungDerEins}
\frac{\m_0}{\sqrt{\pi
t}}\int_{-\frac{\pi}{\mu_0}}^{\frac{\pi}{\mu_0}}\frac{dc}{2\pi}\int_{\R}dp\;e^{-\frac{p^2}{s^2}}\,|c-ip\rangle\langle
c-ip|\;=\;{\rm id}_{\H_{\d}}.
\end{align}

\noindent Furthermore, the coherent states approximate classical
states in the following sense: Given a phase space function
$f(c,\, p)$, the expectation value of the operator $\hat
f\;:=\;f(\hat c,\,\hat p)$ in a coherent state $|c-ip\rangle$ is,
up to corrections in $t$, given by
\begin{align}\label{Gl:ApproximationViaCoherentStates}
\frac{\langle c-ip|\,\hat f\,|c-ip\rangle}{\langle
c-ip|c-ip\rangle}\;\approx\;f(-2c,\,p/s^2),
\end{align}

\noindent where the quality of the approximation depends on $f$,
and will in general not be equally good for all phase space
functions \cite{CCS}.

 To calculate the rigging map corresponding to the Master
constraint $M$ according to (\ref{Gl:RiggingMapEquivalence}), we
need to calculate
\begin{align}\label{Gl:ExponentialExpanded}
\langle \n |e^{-\frac{\hat
M^2}{t}}|\m\rangle\;:=\;\sum_{n=0}^{\infty}\frac{1}{n!}\left(\frac{-1}{t}\right)^n\langle,
\n |\hat M^{2n}|\m\rangle,
\end{align}

\noindent which exists, since the basis vectors can be shown to be analytic vectors for $\hat M$.
Because of (\ref{Gl:ExponentialExpanded}) we need to know matrix elements of even powers of
$\hat M$ in arbitrary basis states $|\n\rangle,\,|\m\rangle$. This
cannot be done analytically, due to the inconvenient structure of
(\ref{Gl:ActionOfVolumeOperator}) in the operator $\hat M$, but
the coherent states will provide a way of approximating $\langle
\n |\hat M^{2n}|\m\rangle$, in the following way:

With $f=M^2$, $M$ being the master constraint of our model, we
will find a generalization of
(\ref{Gl:ApproximationViaCoherentStates}), in particular
\begin{align}
\frac{\langle c'-ip'|\,\hat f\,|c-ip\rangle}{\langle
c'-ip'|c-ip\rangle}\;\approx\;\tilde
f\left(-(c+c')+i(p-p'),\,\frac{(p+p')+i(c-c')}{2s^2}\right),
\end{align}

\noindent where $\tilde f$ is the analytical continuation of $f$
in both variables. This continuation is possible for $f=M^2$,
since $M^2$ is in fact real analytic in both variables. Hence, to
approximate $\langle \n|\hat M^2|\m\rangle$, we first compute the matrix elements of $\hat M^2$ in coherent states
$\langle u|\hat M^2|v\rangle$, with $u=c_1-ip_1$, $v=c_2-ip_2$, and then use
(\ref{Gl:RichtigeZerlegungDerEins}) to write
\begin{align*}
\nbra\,\hat M^2\,\mket\;=\;\frac{\m_0^2}{\pi
t}\int_{[-\frac{\pi}{\mu_0},\frac{\pi}{\mu_0}]^2}\frac{dc_1}{2\pi}\frac{dc_2}{2\pi}\int_{\R^2}dp_1\,dp_2\;
e^{-\frac{p_1^2+p_2^2}{s^2}}\;\nbra
v\rangle\langle v |\,\hat M^2\,|u\rangle\langle u\mket.
\end{align*}

\noindent This will enable us to develop an approximation of
$\langle \n|\hat M^2|\m\rangle$. After that, we will employ the
completeness of the $|\n\rangle$-basis on $\H_{\d}$ to compute
\begin{align}
\langle\n|\hat
M^{2n}|\m\rangle\;=\;\sum_{\l_1,\l_2,\cdots,\l_{n-1}\in\d+4\m_0\Z}\langle\n|\hat
M^2|\l_1\rangle\langle\l_1|\hat
M^2|\l_2\rangle\cdots\langle\l_{n-1}|\hat M^2|\m\rangle.
\end{align}

\noindent Since there is no closed expression for the result
available, we will use another approximation in order to arrive at
a tangible expression for $\langle\n|\hat M^{2n}|\m\rangle$,
the quality of which will depend on $\n,\,\m$, making the
approximation phase-space dependent. Afterwards, we will plug
these approximations for $\langle\n|\hat M^{2n}|\m\rangle$ into
(\ref{Gl:ExponentialExpanded}) and investigate the limit $t\to 0$,
in order to arrive at a final expression for the rigging map
(\ref{Gl:RiggingMapEquivalence}).


\section{Computation of $\langle\n|\hat M^2|\m\rangle$}

\subsection{Expectation value of the Master constraint in coherent states}
\noindent With
\begin{align*}
\sgn\,\hat
p\;\widehat{\sqrt{|p|}}\;|\n\rangle\;&=\;\frac{|\n+\m_0|^{\frac{3}{2}}\,-\,|\n-\m_0|^{\frac{3}{2}}}{3\m_0}\;|\n\rangle\\[5pt]
&=:\;\r_{\n}\,|\n\rangle,
\end{align*}

\noindent the action of the squared Master constraint on a vector
$\nket$ is, via (\ref{Gl:MasterConstraintPowersExpectationValues}) given by
\begin{align}\nonumber
\hat M^2\nket \;=\;\frac{1}{256}\;\Bigg[&\r_{\n}\r_{\n+8\m_0}^2\r_{\n+16\m_0}\;|\n+16\m_0\rangle\\[5pt]\nonumber
\;-4\;&\r_{\n}\r_{\n+12\m_0}\left(\r_{\n+8\m_0}^2\,+\,\r_{\n+4\m_0}^2\right)\;|\n+12\m_0\rangle\\[5pt]\nonumber
\;+\;&\r_{\n}\r_{\n+8\m_0}\left(6\r_{\n+8\m_0}^2\,+\,16\r_{\n+4\m_0}^2\,+\,6\r_{\n}^2\right)\;|\n+8\m_0\rangle\\[5pt]\nonumber
\;-\;&\r_{\n}\r_{\n+4\m_0}\left(4\r_{\n+8\m_0}^2\,+\,24\r_{\n+4\m_0}^2\,+\,24\r_{\n}^2\,+\,4\r_{\n-4\m_0}^2\right)\;|\n+4\m_0\rangle\\[5pt]\nonumber
\;+\;&\r_{\n}^2\left(\r_{\n+8\m_0}^2\,+\,16\r_{\n+4\m_0}^2\,+\,36\r_{\n}^2
\,+\,16\r_{\n-4\m_0}^2\,+\,\r_{\n-8\m_0}^2\right)\;\nket\\[5pt]\nonumber
\;-\;&\r_{\n}\r_{\n-4\m_0}\left(4\r_{\n-8\m_0}^2\,+\,24\r_{\n-4\m_0}^2\,+\,24\r_{\n}^2\,+\,4\r_{\n+4\m_0}^2\right)\;|\n-4\m_0\rangle\\[5pt]\nonumber
\;+\;&\r_{\n}\r_{\n-8\m_0}\left(6\r_{\n-8\m_0}^2\,+\,16\r_{\n-4\m_0}^2\,+\,6\r_{\n}^2\right)\;|\n-8\m_0\rangle\\[5pt]\nonumber
\;-4\;&\r_{\n}\r_{\n-12\m_0}\left(\r_{\n-8\m_0}^2\,+\,\r_{\n-4\m_0}^2\right)\;|\n-12\m_0\rangle\\[5pt]\label{Gl:MQuadrat}
+\;&\r_{\nu}\r_{\n-8\m_0}^2\r_{\n-16\m_0}\;|\n-16\m_0\rangle\Bigg].
\end{align}

\noindent The coherent states in the sector $\d\in[0,4\m_0)$ are
given by
\begin{align*}
|u\rangle\;=\;\sum_{\l\in\d+4\m_0\Z}e^{-\l^2\frac{s^2}{2}}\,e^{i\l
u}\,|\l\rangle.
\end{align*}

\noindent As shown in \cite{PNV04}, this state, labelled by the
complex number $u=c-ip$, is (in our notation) sharply peaked
around the point $(-2c, p/s^2)$ in phase space with width $s$.
One gets, for two coherent states $|u\rangle\,$, $|v\rangle $ :
\begin{align}\label{Gl:Anfangsformel}
&\qquad\langle v | \,\hat M^2\,|u\rangle\;=\;\\[5pt]\nonumber
&\sum_{\text
{terms}}\sum_{\n,\m\in\d+4\m_0\Z}e^{-(\n^2+\m^2)\frac{s^2}{2}}\,e^{i(\n
u -\m\bar v)}\,
\r_{\m+a\m_0}\r_{\m+b\m_0}\r_{\m+c\m_0}\r_{\m+d\m_0}\,\d_{\n,\m+2e\m_0}
\end{align}

\noindent with integer numbers $a,b,\ldots,e$. In order to
evaluate this, we write
\begin{align}\label{Gl:VernachlaessigungR}
\r_{x+a\m_0}\r_{x+b\m_0}\r_{x+c\m_0}\r_{x+d\m_0}\;=\;\left(x\,+\,m'\m_0\right)^2\,+\,\m_0^2\Delta\,+\,R(x)
\end{align}

\noindent where
\begin{align}
m'\,=\,\frac{a+b+c+d}{4},\qquad\Delta\,=\,-\frac{1}{6}\,-\,\frac{1}{16}(a+b+c+d)^2\,-\,\frac{a^2+b^2+c^2+d^2}{4}
\end{align}

\noindent and $\lim_{|x|\to\infty}\,x R(x)<\infty$. If we neglect
$R(x)$, we get
\begin{align}\nonumber
\qquad\langle v | \,\hat M^2\,|u\rangle&\;\approx\;\sum_{\text{terms}}\sum_{\l}e^{-\l^2s^2}\,e^{-(em_0)^2s^2}\,e^{i\l(u-\bar v)}\,e^{i(e\m_0)(u+\bar v)}\,\left[(\l+m\m_0)^2+\mu_0^2\Delta\right]\\[5pt]\label{Gl:NoLabel1}
&\;=\;\sum_{\text {terms}}e^{-\left(\frac{u+\bar
v}{2s^2}\right)^2s^2}\,e^{-\left(\frac{u-\bar v}{2s^2}\right)^2s^2}\,
\,e^{-\left(i\frac{u+\bar v}{2s^2}-em_0\right)^2s^2}\\[5pt]\nonumber
&\qquad\times
\sum_{\l\in\d+4\m_0\Z}e^{-\left(\l-i\frac{u-\bar
v}{2s^2}\right)^2s^2}\,\left[(\l+m\m_0)^2+\mu_0^2\Delta\right],
\end{align}

\noindent with $m=m'-e$. Applying the Poisson summation formula we
get
\begin{align}\nonumber
\sum_{\l\in\d+\m_0\Z}&e^{-\left(\l-i\frac{u-\bar v}{2s^2}\right)^2s^2}\,(\l+m\m_0)^2\\[5pt]
&\;=\;\frac{1}{\m_0}\sum_{n\in\Z}e^{\frac{2\pi n \d
i}{\m_0}}\;\int dx\,e^{-i\frac{2\pi n
}{\m_0}x}\,e^{-\left(x-i\frac{u-\bar
v}{2s^2}\right)^2s^2}\left[(x+m\m_0)^2+\mu_0^2\Delta\right]\\[5pt]\nonumber
&\;=\;\frac{1}{\m_0}\sqrt{\frac{\pi}{s^2}}\;e^{\left(\frac{u-\bar
v}{2s^2}\right)^2s^2}\,e^{-\left(\frac{u-\bar v}{2s^2}-\frac{\pi
n}{\mu_0s^2}\right)^2s^2}\left[\left(i\frac{u-\bar v}{2s^2}-i\frac{\pi
n}{\mu_0s^2}+m\m_0\right)^2\;+\;\frac{1}{2s^2}\;+\;\m_0^2\Delta\right].
\end{align}

\noindent Reinserting this into (\ref{Gl:NoLabel1}) we thus obtain
\begin{align}
\langle v |\,\hat
M^2\,|u\rangle\;&\approx\;\frac{1}{\m_0}\sqrt{\frac{\pi}{s^2}}\sum_{n\in\Z}e^{\frac{2\pi
n \d i}{\m_0}}
e^{-\left(\frac{u-\bar v}{2s^2}-\frac{\pi n}{\m_0 s^2}\right)^2s^2}\\[5pt]\nonumber
&\;\times\;\sum_{\text{terms}} e^{ie\m_0(u+\bar
v)}e^{-(e\m_0)^2s^2}{\scriptstyle\left[\left(i\frac{u-\bar v}{2s^2}-i
\frac{\pi n}{\m_0
s^2}\right)^2+\frac{1}{s^2}\left(\frac{1}{2}+2i\left(\frac{u-\bar
v}{2s^2}-\frac{\pi n}{\m_0 s^2}\right)m\m_0
\right)+\m_0^2(m^2+\Delta)\right]}.
\end{align}

\noindent Calculating the sum over the different terms explicitly,
taking the prefactors in (\ref{Gl:MQuadrat}) into account, we
obtain
\begin{align*}
&\sum_{\text{terms}}  e^{ie\m_0(u+\bar
v)}e^{-(e\m_0)^2s^2}\;=\;e^{s^2\frac{\del^2}{\del z^2}}{}
_{\big|z=i(u+\bar v){\m_0}}\;\sin^8z\\[5pt]
&\sum_{\text{terms}} e^{ie\m_0(u+\bar v)}e^{-(e\m_0)^2s^2}\; m \;=\;0\\[5pt]
&\sum_{\text{terms}}e^{ie\m_0(u+\bar
v)}e^{-(e\m_0)^2s^2}\;(m^2\,+\,\Delta)\\[5pt]
 &\qquad\qquad
\;=\;\frac{1}{16}\;e^{s^2\frac{\del^2}{\del z^2}}{}_{\big|z=i(u+\bar
v){\m_0}}\;\left(
-\frac{193}{48}\cos\;8\m_0z\;+\frac{109}{6}\cos\;6\m_0z\right.\\[5pt]\nonumber
&\qquad\qquad\qquad\left.\;-\;\frac{487}{12}\cos\;4\m_0z\;+\frac{55}{6}\cos\;2\m_0z\;-\frac{35}{48}
\right)\;\\[5pt]
&\qquad\qquad\;
 =:\;\frac{1}{16}\;e^{s^2\frac{\del^2}{\del z^2}}{}_{\big|z=i(u+\bar
v){\m_0}}\;\z(z)
\end{align*}

\noindent So
\begin{align}\label{Gl:ErsteRichtigeZwischenFormel}
\langle v |\,\hat
M^2\,|u\rangle\;&\approx\;\frac{1}{\m_0}\sqrt{\frac{\pi}{s^2}}\sum_{n\in\Z}e^{\frac{2\pi
n \d i}{\m_0}}e^{-\left(\frac{u-\bar v}{2s^2}-\frac{\pi n}{\m_0
s^2}\right)^2s^2}\\[5pt]\nonumber
&\;\times\;\left[e^{s^2\frac{\del^2}{\del z^2}}{}_{\big|z=i(u+\bar
v)\m_0}\sin^8z\;\left({\scriptstyle\left(i\frac{u-\bar
v}{2s^2}-i\frac{\pi n}{\m_0
s^2}\right)^2\;+\;\frac{1}{2s^2}}\right)\;+\;\frac{\m_0^2}{16}e^{s^2\frac{\del^2}{\del
z^2}}{}_{\big|z=i(u+\bar v)\m_0}\z(z)\right].
\end{align}


\subsection{Matrix elements of $\hat M^2$ with respect to the basis vectors}
\noindent We use the result from the last section to calculate now
arbitrary matrix elements of $\hat M^2$. For this, we need the
resolution of unity provided by the coherent states (\ref{Gl:RichtigeZerlegungDerEins}):
\begin{align}
\frac{\m_0}{\sqrt{\pi
s^2}}\int_{-\frac{\pi}{\mu_0}}^{\frac{\pi}{\mu_0}}\frac{dc}{2\pi}\int_{\R}dp\;e^{-\frac{p^2}{s^2}}\,|c-ip\rangle\langle
c-ip|\;=\;\text{id}_{\H_{\d}}.
\end{align}

\noindent so we have:
\begin{align}\label{Gl:ZerlegungDerEins}
\nbra\,\hat M^2\,\mket\;=\;\frac{\m_0^2}{\pi
s^2}\int_{[-\frac{\pi}{\mu_0},\frac{\pi}{\mu_0}]^2}\frac{dc_1}{2\pi}\frac{dc_2}{2\pi}\int_{\R^2}dp_1\,dp_2\;e^{-\frac{p_1^2+p_2^2}{s^2}}\;\nbra
v\rangle\langle v |\,\hat M^2\,|u\rangle\langle u\mket
\end{align}

\noindent with $u=c_2-ip_2$, $v=c_1-ip_1$. To make life simpler,
we will split (\ref{Gl:ErsteRichtigeZwischenFormel}) up into
several parts before inserting and perform the integrations
separately. Expanding the first exponential derivative in
(\ref{Gl:ErsteRichtigeZwischenFormel}) into a power series, we
obtain
\begin{align}\nonumber
\langle v |\,\hat M^2\,| u \rangle
&\;=\;\frac{1}{\m_0}\sqrt{\frac{\pi}{s^2}}\sum_{n\in\Z}e^{\frac{2\pi
n \d i}{\m_0}}e^{-\left(\frac{u-\bar v}{2s^2}-\frac{\pi n}{\m_0
s^2}\right)^2s^2}\;\sin^{8}\left(\m_0(u+\bar
v)\right)\;\left({\scriptstyle\left(i\frac{u-\bar
v}{2s^2}-i\frac{\pi n}{\m_0
s^2}\right)^2\;+\;\frac{1}{2s^2}}\right)\\[5pt]\label{Gl:FetteZerlegung}
&\;+\;\frac{1}{\m_0}\sqrt{\frac{\pi}{s^2}}\,\sum_{m=1}^{\infty}\frac{s^{2m}}{m!}\frac{\del^{2m}}{\del
z^{2m}}_{\big|z=\left(\m_0(u+\bar
v)\right)}\;\sin^{8}\,z\;\\[5pt]\nonumber
&\qquad\qquad\times\sum_{n\in\Z}e^{\frac{2\pi n \d
i}{\m_0}}e^{-\left(\frac{u-\bar v}{2s^2}-\frac{\pi n}{\m_0
s^2}\right)^2s^2}\;\left({\scriptstyle\left(i\frac{u-\bar
v}{2s^2}-i\frac{\pi n}{\m_0
s^2}\right)^2\;+\;\frac{1}{2s^2}}\right)\\[5pt]\nonumber
&\;+\;\frac{1}{\m_0}\sqrt{\frac{\pi}{s^2}}\sum_{n\in\Z}e^{\frac{2\pi
n \d i}{\m_0}}e^{-\left(\frac{u-\bar v}{2s^2}-\frac{\pi n}{\m_0
s^2}\right)^2s^2}\;\frac{\m_0^2}{16}e^{s^2\frac{\del^2}{\del
z^2}}{}_{\big|z=i(u+\bar v)\m_0}\;\z(z).
\end{align}

\noindent We insert now term by term of (\ref{Gl:FetteZerlegung})
into (\ref{Gl:ZerlegungDerEins}). We start with the zeroth order
term of the derivative expansion:
\begin{align}\nonumber
\frac{\m_0}{\sqrt{\pi
s^6}}\int_{[-\frac{\pi}{\mu_0},\frac{\pi}{\mu_0}]^2}\frac{dc_1}{2\pi}&\frac{dc_2}{2\pi}
\int_{\R^2}dp_1\,dp_2\;e^{-\left(\n-\frac{p_1}{s^2}\right)^2\frac{s^2}{2}}\;e^{-\left(\m-\frac{p_2}{s^2}\right)^2\frac{s^2}{2}}
\;e^{i\n c_1}\,e^{-i\m
c_2}\,e^{-\frac{p_1^2+p_2^2}{2s^2}}\\[5pt]
&\times\;\sin^{8}\Big[{\m_0(c_1+c_2-i(p_2-p_1))}\Big]\\[5pt]\nonumber
&\times\;\sum_{n\in\Z}e^{\frac{2\pi i n
\d}{\m_0}}\;e^{\left(\frac{p_2+p_1-i(c_2-c_1)}{2s^2}+i\frac{\pi
n}{\m_0 s^2}\right)^2s^2}\\[5pt]\nonumber
&\qquad\times
\left[\left(\frac{p_2+p_1-i(c_2-c_1)}{2s^2}+i\frac{\pi n}{\m_0
s^2}\right)^2\;+\;\frac{1}{2s^2}\right].
\end{align}

\noindent Since the exponentials and the sine function are all
invariant under the substitution $c_1\to c_1+\frac{2 \pi}{\m_0}$,
this integral has the following form:
\begin{align}
\int_{-\frac{\pi}{\m_0}}^{-\frac{\pi}{\m_0}}\frac{dc_1}{2\pi}\;\sum_{n=0}^{\infty}\;f\left(c_1+{\frac{2\pi
n}{\m_0}} \right)\,e^{\frac{2\pi i n
\d}{\m_0}}\;=\;\int_{\R}\frac{dc_1}{2\pi}\;f(c_1)\;\c(c_1),
\end{align}

\noindent where
\begin{align}
\c(c_1)\;:=\;\sum_{n=0}^{\infty}\;\c_{[-\frac{\pi}{\mu_0},\frac{\pi}{\mu_0}]}(c_1)\;e^{\frac{2\pi
i n \d}{\m_0}}
\end{align}

\noindent is a sum over characteristic functions. This gives us
\begin{align}\nonumber
\frac{\m_0}{\sqrt{\pi
s^6}}\int_{\R}&\frac{dc_1}{2\pi}\int_{-\frac{\pi}{\mu_0}}^{\frac{\pi}{\mu_0}}\frac{dc_2}{2\pi}
\int_{\R^2}dp_1\,dp_2\;e^{-\left(\n-\frac{p_1}{s^2}\right)^2\frac{s^2}{2}}\;e^{-\left(\m-\frac{p_2}{s^2}\right)^2\frac{s^2}{2}}
\;e^{i\n c_1}\,e^{-i\m c_2}\,e^{-\frac{p_1^2+p_2^2}{2s^2}}\\[5pt]
&\times\;\sin^{8}\Big[{ \m_0(c_1+c_2-i(p_2-p_1))}\Big]\\[5pt]\nonumber
&\times\;\c(c_1)\;e^{\left(\frac{p_2+p_1-i(c_2-c_1)}{2s^2}\right)^2s^2}
\left[\left(\frac{p_2+p_1-i(c_2-c_1)}{2s^2}\right)^2\;+\;\frac{1}{2s^2}\right].
\end{align}

\noindent Now consider the integration over $c_1$. Notice that the
modulus of the integrand is dominated by a Gaussian times a
polynomial, where the width of the Gaussian is proportional to
$s$. So, for small semiclassicality parameter $s$, the main
contribution of the integral over $c_1$ will come from the point
where the gaussian is peaked, that is
$-c_2\in[-\frac{\pi}{\m_0},\frac{\pi}{\m_0}]$. At this point,
$\c(-c_2)=1$. Therefore we have for the integral:

\begin{align}\nonumber
\frac{\m_0}{\sqrt{\pi
s^6}}\int_{\R}&\frac{dc_1}{2\pi}\int_{-\frac{\pi}{\mu_0}}^{\frac{\pi}{\mu_0}}\frac{dc_2}{2\pi}
\int_{\R^2}dp_1\,dp_2\;e^{-\left(\n-\frac{p_1}{s^2}\right)^2\frac{s^2}{2}}\;e^{-\left(\m-\frac{p_2}{s^2}\right)^2\frac{s^2}{2}}
\;e^{i\n c_1}\,e^{-i\m c_2}\,e^{-\frac{p_1^2+p_2^2}{2s^2}}\\[5pt]\nonumber
&\times\;\sin^{8}\Big[{\m_0(c_1+c_2-i(p_2-p_1))}\Big]\\[5pt]
&\times\;e^{\left(\frac{p_2+p_1-i(c_2-c_1)}{2s^2}\right)^2s^2}
\left[\left(\frac{p_2+p_1-i(c_2-c_1)}{2s^2}\right)^2\;+\;\frac{1}{2s^2}\right]\\[5pt]\nonumber
&\times\;(1+O(s^{\infty}))
\end{align}

\noindent We omit the $(1+O(s^{\infty}))$ in the following
calculations. Shifting the integration of $c_1$ by $c_2$ and
rescaling the integration of $c_2$ by $\m_0$, we obtain
\begin{align}\nonumber
\frac{1}{\sqrt{\pi
s^6}}\int_{\R}&\frac{dc_1}{2\pi}\int_{-\pi}^{\pi}\frac{dc_2}{2\pi}\int_{\R^2}dp_1\,dp_2\;e^{-\left(\n-\frac{p_1}{s^2}
\right)^2\frac{s^2}{2}}\;e^{-\left(\m-\frac{p_2}{s^2}\right)^2\frac{s^2}{2}}\;e^{i\n
c_1-i\frac{\m-\n}{\m_0}c_2}\,e^{-\frac{p_1^2+p_2^2}{2s^2}}\\[5pt]\label{Gl:PunktAnDemOVonTUnendlich}
&\times\;\sin^{8}\Big[{ 2c_2+\m_0(c_1-i(p_2-p_1))}\Big]\\[5pt]\nonumber
&\times\;e^{-\left(\frac{c_1-i(p_1+p_2)}{2s^2}\right)^2s^2}
\left[\frac{1}{2s^2}\;-\;\left(\frac{c_1-i(p_1+p_2)}{2s^2}\right)^2\right]
\end{align}

\noindent Note that the integrand is analytic in $c_1$. We now
make a shift of the contour of integration of $c_2$ by
$i(p_1+p_2)$. When doing this, one has to take care about the
convergence of the integral. As one can check, no problem arises in our case. We
obtain
\begin{align}\nonumber
\frac{1}{\sqrt{\pi
s^6}}\int_{\R}&\frac{dc_1}{2\pi}\int_{-\pi}^{\pi}\frac{dc_2}{2\pi}\int_{\R^2}dp_1\,dp_2\;e^{-\left(\n-\frac{p_1}{s^2}
\right)^2\frac{s^2}{2}}\;e^{-\left(\m-\frac{p_2}{s^2}\right)^2\frac{s^2}{2}}\;e^{\n(p_1+p_2)}\;e^{i\n
c_1-i\frac{\m-\n}{\m_0}
c_2}\,e^{-\frac{p_1^2+p_2^2}{2s^2}}\\[5pt]
&\times\;\sin^{8}\big({
2c_2+\m_0(c_1+2ip_2)}\big)\;e^{-\frac{c_1^2}{4s^2}}
\left(\frac{1}{2s^2}\;-\;\frac{c_1^2}{4s^4}\right)
\end{align}

\noindent It is now convenient to carry out the integration over
$c_2$. We notice that
\begin{align}
\int_{-\pi}^{\pi}\frac{dc_2}{2\pi}\;\sin^8(2c_2+a)\;e^{iNc_2}\;=\;\frac{1}{256}(-1)^{\frac{N}{4}}\left(\begin{array}{c}8
\\4+{\scriptstyle\frac{N}{4}}\end{array}\right)\,e^{-ia\frac{N}{2}},
\end{align}

\noindent when $N$ is divisible by four.  We get:
\begin{align}\nonumber
\frac{1}{256}(-1)^{\NN}\frac{1}{\sqrt{\pi
s^6}}\left(\begin{array}{c}8\\4+{\scriptstyle\NN}
\end{array}\right)\int_{\R^3}&\frac{dc_1}{2\pi}dp_1\,dp_2\;e^{-\left(\n-\frac{p_1}{s^2}\right)^2\frac{s^2}{2}}\;
e^{-\left(\m-\frac{p_2}{s^2}\right)^2\frac{s^2}{2}}\;e^{\n p_2+\m p_1}\;e^{-\frac{p_1^2+p_2^2}{2s^2}}\\[5pt]
&\times\; \;e^{-i\frac{\n+\m}{2} c_1}\;e^{-\frac{c_1^2}{4s^2}}
\left(\frac{1}{2s^2}\;-\;\frac{c_1^2}{4s^4}\right).
\end{align}

\noindent Integration over $c_1$ gives:
\begin{align}\nonumber
\frac{1}{256}(-1)^{\NN}\frac{1}{\pi s^2}\left(\begin{array}{c}8\\4+{\scriptstyle\NN}\end{array}\right)\int_{\R^2}&dp_1\,dp_2\;e^{-\left(\n-\frac{p_1}{s^2}\right)^2\frac{s^2}{2}}\;e^{-\left(\m-\frac{p_2}{s^2}\right)^2\frac{s^2}{2}}\;e^{\n p_2+\m p_1}\;e^{-\frac{p_1^2+p_2^2}{2s^2}}\\[5pt]
&\times\;
e^{-s^2\left(\frac{\n+\m}{2}\right)^2}\;\left(\frac{\n+\m}{2}\right)^2\\[5pt]\nonumber
&\;=\;\frac{1}{256}(-1)^{\NN}\left(\begin{array}{c}8\\4+{\scriptstyle\NN}\end{array}\right)\left(\frac{\n+\m}{2}\right)^2\;e^{\n\m
s^2}.
\end{align}

\noindent We now reinsert the correction that we have omitted
since formula (\ref{Gl:PunktAnDemOVonTUnendlich}) and are left
with

\begin{align}
\frac{1}{256}(-1)^{\NN}\left(\begin{array}{c}8\\4+{\scriptstyle\NN}\end{array}\right)\left(\frac{\n+\m}{2}\right)^2\;
e^{\n\m s^2}\;(1+O(s^{\infty})).
\end{align}

\noindent For all higher orders of the exponentiated derivative in
(\ref{Gl:FetteZerlegung}) we obtain similar but more complicated
expressions, that have in common the fact that they stay finite,
as $s$ tends to $0$. The integration over the term containing $\z$
in (\ref{Gl:FetteZerlegung}) is easily performed and yields, up to
terms of order $O(s^2)$:
\begin{align}\nonumber
\frac{\m_0^3}{16\sqrt{\pi
s^6}}\int_{[-\frac{\pi}{\mu_0},\frac{\pi}{\mu_0}]^2}\frac{dc_1}{2\pi}&\frac{dc_2}{2\pi}
\int_{\R^2}dp_1\,dp_2\;e^{-\left(\n-\frac{p_1}
{s^2}\right)^2\frac{s^2}{2}}\;e^{-\left(\m-\frac{p_2}{s^2}\right)^2\frac{s^2}{2}}\;\\[5pt]\nonumber
&\qquad\; \times \;e^{i\n c_1}\,e^{-i\m
c_2}\,e^{-\frac{p_1^2+p_2^2}{2s^2}}\; e^{s^2\frac{\del^2}{\del
z^2}}{}_{\big|z=i(u+\bar v)\m_0}\;\z(z)\\[5pt]
&\times\;\sum_{n\in\Z}e^{\frac{2\pi i n
\d}{\m_0}}\;e^{\left(\frac{p_2+p_1-i(c_2-c_1)}{2s^2}+i\frac{\pi
n}{\m_0 s^2}\right)^2s^2}\\[5pt]\nonumber
&\;=\;\frac{\m_0^2}{32}\left[\frac{193}{48}\,\d_{\left|\NN\right|,4}\;+\;\frac{109}{6}\,
\d_{\left|\NN\right|,3}\right.\\[5pt]\nonumber
&\qquad\;+\;\left.\frac{487}{12}\,\d_{\left|\NN\right|,2}\;+\frac{55}{6}\,
\d_{\left|\NN\right|,1}\;+\;\frac{35}{48}\,\d_{\left|\NN\right|,0}\right]\;(1+O(s^2)).
\end{align}

\noindent We now go back to the formula for the resolution of
unity in (\ref{Gl:RichtigeZerlegungDerEins}) and notice that the
right hand side does not depend on $s$, so
\begin{align}
\text{id}_{\H_{\d}}&\;=\;\lim_{s\to 0}\;\text{id}_{\H_{\d}}\\[5pt]\nonumber
&\;=\;\lim_{s\to 0}\quad\frac{\m_0}{\sqrt{\pi
s^2}}\int_{-\frac{\pi}{\mu_0}}^{\frac{\pi}{\mu_0}}\frac{dc}{2\pi}\int_{\R}dp\;e^{-\frac{p^2}{s^2}}\,|c-ip\rangle\langle
c-ip|.
\end{align}

\noindent This allows us to drop all terms of order $O(s)$ in our
approximation, and we are left with
\begin{align}\nonumber
\nbra\,\hat
M^2\,\mket\;\approx\;\frac{1}{256}&(-1)^{\NN}\left(\begin{array}{c}8\\4+
{\scriptstyle\NN}\end{array}\right)\left(\frac{\n+\m}{2}\right)^2\\[5pt]
&\;-\;\frac{\m_0^2}{32}(-1)^{\NN}\left[\frac{193}{48}\,\d_{\left|\NN\right|,4}\;+\;\frac{109}{6}\,\d_{\left|\NN\right|,3}\right.\\[5pt]\nonumber
&\qquad\;+\;\left.\frac{487}{12}\,\d_{\left|\NN\right|,2}\;+\frac{55}{6}\,\d_{\left|\NN\right|,1}\;+\;\frac{35}{48}\,\d_{\left|\NN\right|,0}\right].
\end{align}

\noindent Note that the approximation now comes from the
neglection of the $R(x)$ - term in formula
(\ref{Gl:VernachlaessigungR}), but not any more from
$s$-corrections, since we have gone over to the limit $s\to 0$, in
which all these terms vanish.


\section{Computation of $\langle\n|\hat M^{2n}|\m\rangle$ for $n>1$ and rigging map}
\subsection{Expectation values of $\hat M^{2n}$}
\noindent The approximate formula for the matrix elements of $\hat
M^2$ have been calculated in the last section. As demanded the
approximation vanishes if $\NN$ is not an integer. If it is, the
matrix elements can be approximated by
\begin{align*}
\nbra\,\hat M^2\,\mket\;\approx\;\\[5pt]
\frac{(-1)^N}{256}&
\Bigg[\left(\begin{array}{c}8\\4+N\end{array}\right)\left(\frac{\n+\m}{2}\right)^2\\[5pt]\nonumber
&\;+\;\frac{193}{6}\m_0^2\d_{|N|,4}\;+\;\frac{436}{3}\m_0^2\d_{|N|,3}\;+\;\frac{974}{3}\m_0^2\d_{|N|,2}\;+\;\frac{220}{3}\m_0^2\d_{|N|,1}\;+\;\frac{35}{6}\m_0^2\d_{|N|,0}\Bigg]
\end{align*}

\noindent with $N=\NN$. So
\begin{align}\label{Gl:ErsteGleichung}
\nbra\,\hat
M^2\,\mket\;=\;\frac{(-1)^N}{256}\left(\begin{array}{c}8\\4+N\end{array}\right)\left(\frac{\n+\m}{2}\right)^2\;+
\;R(\n,\m,n),
\end{align}

\noindent with $R$ being correction terms that we neglect for the
time being.\\

\noindent Since the $\nket$ with $\n\in\d+4\m_0\Z$ provide a basis
of $\H_{\d}$, we can easily compute the matrix elements of $\hat
M^{2n}$:
\begin{align*}
\nbra\,\hat M^{2n}\,\mket\;\approx\;\frac{(-1)^N}{256^n}\sum_{\scriptsize\begin{array}{c}\l_1\in\d+4\m_0\Z\\\vdots\\
\l_{n-1}\in\d+4\m_0\Z\end{array}}&\left(\begin{array}{c}8\\4+\frac{\n-\l_1}{4\m_0}\end{array}\right)\left(
\begin{array}{c}8\\4+\frac{\l_1-\l_2}{4\m_0}\end{array}\right)\cdots\left(\begin{array}{c}8\\4+\frac{\l_{n-1}-\m}{4\m_0}
\end{array}\right)\\[5pt]
&\times\left(\frac{\n+\l_1}{2}\right)^2\left(\frac{\l_1+\l_2}{2}\right)^2\cdots\left(\frac{\l_{n-1}+\m}{2}\right)^2.
\end{align*}

\noindent It is very difficult to give a closed expression for
this formula. Still, we want to show a way of calculating these
matrix elements in principle. For this, we first rewrite the
summation:
\begin{align}\label{Gl:AusgangsGleichungMitSumme}
\nbra\,\hat M^{2n}\,\mket\;\approx\;\frac{(-1)^{\NN}}{256^n}\sum_{\scriptsize\begin{array}{c}l_1,\ldots,\l_n\in\Z\\
\sum_k
l_k=\frac{\n-\m}{4\m_0}\end{array}}&\prod_{k=1}^n\left[\left(\begin{array}{c}8\\4+l_k\end{array}\right)\right]
\\[5pt]\nonumber
\times&\prod_{k=1}^n\left[\left(\frac{\n+\m}{2}\;-\;\m_0\sum_{j=1}^{k-1}l_j\;+\;\m_0\sum_{j=k+1}^nl_j\right)^2\right].
\end{align}

\noindent Note that, whenever $\NN$ is not an integer, the sum is
empty and thus, by convention, $0$, as demanded. Because of
\begin{align}
\left(\frac{\n+\m}{2}\;-\;\m_0\sum_{j=1}^{k-1}l_j\;+\;\m_0\sum_{j=k+1}^nl_j\right)^2\;=\;
\frac{d^2}{da_k^2}_{\Big|a_k=0}\,e^{a_k\left(\frac{\n+\m}{2}\;-\;\m_0\sum_{j=1}^{k-1}l_j\;+\;\m_0\sum_{j=k+1}^nl_j\right)}
\end{align}

\noindent and
\begin{align}
\d_{\sum_kl_k,\,\NN}\;=\;\int_0^{2\pi}\frac{db}{2\pi}e^{ib\left(\NN-\sum_kl_k\right)},
\end{align}

\noindent We can rewrite (\ref{Gl:AusgangsGleichungMitSumme}) in
the following way, after shifting each of the sums over $l_k$ by
$-4$:
\begin{align}\nonumber
\nbra\,\hat
M^{2n}\,\mket\;\approx\;\frac{(-1)^{\NN}}{256^n}&\sum_{l_1,\ldots,l_n\in\Z}\int_0^{2\pi}\frac{db}{2\pi}
e^{ib\left(\NN-\sum_kl_k+4n \right)}\;
\left[\prod_{k=1}^n\left(\begin{array}{c}8\\l_k\end{array}\right)\right]
\\[5pt]\nonumber
\times\;&\prod_{k=1}^n\left[\frac{d^2}{da_k^2}_{\Big|a_k=0}\,e^{a_k\left(\frac{\n+\m}{2}\;-\;\m_0\sum_{j=1}^{k-1}l_j\;+\;\m_0
\sum_{j=k+1}^nl_j\;+\;4\mu_0(n-2k+1)\right)}\right]\\[5pt]\label{Gl:NochEineGleichung}
\;=\;\frac{(-1)^{\NN}}{256^n}&\sum_{l_1,\ldots,l_n\in\Z}\int_0^{2\pi}\frac{db}{2\pi}e^{ib\left(\NN-\sum_kl_k+4n
\right)}
\;\left[\prod_{k=1}^n\left(\begin{array}{c}8\\l_k\end{array}\right)\right]\\[5pt]\nonumber
\times\;&\left[\prod_{k=1}^n\frac{d^2}{da_k^2}_{\Big|a_k=0}\,\right]\prod_{k=1}^n\left[\frac{d^2}{da_k^2}_{\Big|a_k=0}\,
e^{4a_k\mu_0(n-2k+1)}\;e^{l_k\m_0\left(\sum_{j=1}^{k-1}a_j\;-\;\sum_{j=k+1}^na_j\right)}\right]
\end{align}

\noindent By the binomial formula, this is
\begin{align}\nonumber
\nbra\,\hat
M^{2n}\,\mket\;&\approx\;\frac{(-1)^{\NN}}{256^n}\int_0^{2\pi}\frac{db}{2\pi}\;e^{ib\left(\NN+4n
\right)}
\\[5pt]
&\times\;\left[\prod_{k=1}^n\frac{d^2}{da_k^2}_{\Big|a_k=0}\,\right]\;\prod_{k=1}^n
\,e^{4a_k\mu_0(n-2k+1)}\, \left(1+\z_k(a_1,\ldots,a_n,b)\right)^8
\end{align}
\noindent with
\begin{align}
\z_k(a_1,\ldots,a_n,b)\;=\;e^{\m_0\left(\sum_{j=1}^{k-1}a_j-\sum_{j=k+1}^na_j\right)\,-ib}.
\end{align}

\noindent It is now straightforward, though tedious, do perform
the $2n$ differentiations. Since
\begin{align}
\frac{d}{da_k}(1+\z_l(\vec
a,b))^n\;=\;n\,\sgn(l-k)\left((1+\z_l(\vec a,b))^n-(1+\z_l(\vec
a,b))^{n-1}\right),
\end{align}

\noindent it is predictable that the differentiations and setting
of $\vec a=0$ afterwards will yield the result
\begin{align}
\left[\prod_{k=1}^n\frac{d^2}{da_k^2}_{\Big|a_k=0}\,\right]\;&\prod_{k=1}^n
\,e^{4a_k\mu_0(n-2k+1)}\,
\left(1+\z_k(a_1,\ldots,a_n,b)\right)^8\\[5pt]\nonumber
&\;=\;\sum_{m=0}^{8n}P_{n,m}(\n,\m)(1+e^{-ib})^{8n-m}e^{ibm},
\end{align}

\noindent where the $P_{n,m}(\n,\m)$ are polynomials in $\n$ and
$\m$. Since
\begin{align}
\int_0^{2\pi}\frac{db}{2\pi}e^{ib\left(\NN+4n
\right)}(1+e^{-ib})^{8n-m}e^{ibm}\;=\;\left(\begin{array}{c}8n-m\\4n+\NN-m\end{array}\right)
\end{align}

\noindent and
\begin{align}\label{Gl:RelateBinomialCoefficients}
\left(\begin{array}{c}8n-m\\4n+\NN-m\end{array}\right)\;&=\;\left(\frac{4n+\NN}{8n}\right)\left(\frac{4n+\NN-1}{8n-1}
\right)\cdots\\[5pt]\nonumber
&\qquad\qquad\cdots\left(\frac{4n+\NN-m+1}{8n-m+1}\right)\left(\begin{array}{c}8n\\4n+\NN\end{array}\right),
\end{align}

\noindent the form of the result is known:
\begin{align}
\nbra\,\hat
M^{2n}\,\mket\;\approx\;\frac{(-1)^{\NN}}{256^n}\;Q_n(\n,\m)\;\left(\begin{array}{c}8n\\4n+\NN\end{array}\right),
\end{align}

\noindent where $Q_n(\n,\m)$ is another polynomial in $\n$ and
$\m$, consisting of linear combinations of the $P_{n,m}(\n,\m)$.
However, it is a considerable computational task to obtain the
polynomials $P_{n,m}(\n,\m)$ for higher $m$, and, to the best of
the authors' knowledge, no closed formula for them exists. So,
although one can compute the approximate matrix elements of
$\nbra\,\hat M^{2n}\,\mket$ explicitly order by order, no closed
formula for them is known.


\subsection{Further approximations}\label{Ch:FurtherApproximations}

\noindent In order to compute the rigging map approximatively, we
start again from formula (\ref{Gl:AusgangsGleichungMitSumme})
\begin{align}\label{Gl:AusgangsGleichungMitSumme2}
\nbra\,\hat
M^{2n}\,\mket\;\approx\;\frac{(-1)^{\NN}}{256^n}\sum_{\scriptsize\begin{array}
{c}l_1,\ldots,l_n\in\Z\\\sum_k
l_k=\frac{\n-\m}{4\m_0}\end{array}}&\prod_{k=1}^n\left[\left(
\begin{array}{c}8\\4+l_k\end{array}\right)\right]\\[5pt]\nonumber
\times&\prod_{k=1}^n\left[\left(\frac{\n+\m}{2}\;-\Delta_k(\vec
l)\;\right)^2\right].
\end{align}

\noindent with
\begin{align}
\Delta_k(\vec
l)\;:=\;\m_0\sum_{j=1}^{k-1}l_j\;-\;\m_0\sum_{j=k+1}^nl_j.
\end{align}

\noindent Now assume that $\nu-\mu$ is small compared to $\n+\m$.
Then the summation of the $l_k$ ranges over $-4,\ldots,4$, and
since the binomial prefactors are nearly Gaussian distributions in
the $l_k$, it is safe to assume that the largest contribution to the
sum will be made by products, where all the $l_k$ are close to 0,
compared to $4$, hence the $\Delta_k(\vec l)$ are much smaller
that $\frac{\n+\m}{2}$. So, as a test, we could expand formula
(\ref{Gl:AusgangsGleichungMitSumme2}) into powers of the sums over
the $\Delta_k(\vec l)$:

\begin{align}\nonumber
\nbra\,\hat
M^{2n}\,\mket\;&\approx\;\frac{(-1)^{\NN}}{256^n}\sum_{\scriptsize\begin{array}{c}l_1,
\ldots,l_n\in\Z\\\sum_k
l_k=\frac{\n-\m}{4\m_0}\end{array}}\prod_{k=1}^n\left[\left(\begin{array}
{c}8\\4+l_k\end{array}\right)\right]\\[5pt]\label{Gl:EntwicklungInDeltaK}
&\;\times\;\left[\left(\frac{\n+\m}{2}\right)^{2n}\;-\;2\sum_{k=1}^n\Delta_k(\vec
l)\left(\frac{\n+\m}{2}\right)^{2n-1}\right.\\[5pt]\nonumber
&\quad\;\left. +\;\left(2\left(\sum_{k=1}^n\Delta_k(\vec
l)\right)^2\,-\,\sum_{k=1}^{n}\Delta_k(\vec
l)^2\right)\left(\frac{\n+\m}{2}\right)^{2n-2}\;+\;\ldots\right].
\end{align}

\noindent For what follows, we assume $n>1$, since only then are
the $\Delta_k(\vec l)\neq 0$. The cases $n=0$ and $n=1$ are
trivial and already known respectively.

 Now first we observe that
\begin{align}
\sum_{\scriptsize\begin{array}{c}l_1,\ldots,l_n\in\Z\\\sum_k
l_k=\frac{\n-\m}{4\m_0}\end{array}}f(\vec
l)\;=\;\sum_{\scriptsize\begin{array}{c}l_1,\ldots,l_n\in\Z\\\sum_k
l_k=\frac{\n-\m}{4\m_0}\end{array}}\frac{1}{n!}\sum_{\s\in\Sym_n}f(\s(\vec
l)),
\end{align}

\noindent where $\Sym_n$ is the permutation group in $n$ elements
and $\s(\vec l)$ means the vector, of which the $n$ entries in the
vector $\vec l$ are permuted according to $\s$. Our second
observation is that
\begin{align}
\sum_{\s\in\Sym_n} \sum_{k=1}^n\Delta_k(\s(\vec l))\;=\;0,
\end{align}

\noindent so the "linear" term in (\ref{Gl:EntwicklungInDeltaK})
vanishes, and, keeping only the first two terms of our expansion,
we are left with
\begin{align}\label{Gl:StepWhereFurtherApproximationIsMade}
\nbra\,\hat
M^{2n}\,\mket\;&\approx\;\frac{(-1)^{\NN}}{256^n}\sum_{\scriptsize\begin{array}{c}
l_1,\ldots,l_n\in\Z\\\sum_k
l_k=\frac{\n-\m}{4\m_0}\end{array}}\prod_{k=1}^n\left[\left(
\begin{array}{c}8\\4+l_k\end{array}\right)\right]\\[5pt]\nonumber
&\quad\;\times\left[\left(\frac{\n+\m}{2}\right)^{2n}\;+\;\left(2\left(\sum_{k=1}^n\Delta_k(\vec
l)\right)^2\,-\,\sum_{k=1}^{n}\Delta_k(\vec
l)^2\right)\left(\frac{\n+\m}{2}\right)^{2n-2}\right]
\end{align}
\begin{align}\nonumber
\;&=\;\frac{(-1)^{\NN}}{256^n}\left(\begin{array}{c}8n\\4n+\NN\end{array}\right)\left(\frac{\n+\m}{2}\right)^{2n}
\\[5pt]\nonumber
\;&\quad+\;\frac{(-1)^{\NN}}{256^nn!}\left(\frac{\n+\m}{2}\right)^{2n-2}\\[5pt]\label{Gl:ExpansionCarriedOut}
&\quad\quad\times\left[\;2\sum_{\scriptsize\begin{array}{c}l_1,\ldots,l_n\in\Z\\\sum_k
l_k=\frac{\n-\m}{4\m_0}\end{array}}\prod_{k=1}^n\left[\left(\begin{array}{c}8\\4+l_k\end{array}\right)\right]
\sum_{\s\in\Sym_n}\left(\sum_{k=1}^n\Delta_k(\s(\vec
l))\right)^2\right.\\[5pt]\nonumber
&\left.\quad\quad-\;\sum_{\scriptsize\begin{array}{c}l_1,\ldots,l_n\in\Z\\\sum_k
l_k=\frac{\n-\m}{4\m_0}\end{array}}\prod_{k=1}^n\left[\left(\begin{array}{c}8\\4+l_k\end{array}\right)\right]
\sum_{\s\in\Sym_n}\sum_{k=1}^n\Delta_k(\s(\vec l))^2\right]
\end{align}

\noindent Let us consider the middle term in formula
(\ref{Gl:ExpansionCarriedOut}). First we note that
\begin{align}\label{Gl:LustigeZahlenreihe}
\frac{1}{\m_0}\sum_{k=1}^n\Delta_k(\vec
l)\;=\;\left\{\begin{array}{ll}(n-1)(l_n-l_1)+(n-3)(l_{n-1}-l_2)+
\cdots+2(l_{\frac{n+3}{2}}-l_{\frac{n-1}{2}})&n \text{ odd}\\[5pt](n-1)(l_n-l_1)+(n-3)(l_{n-1}-l_2)+\cdots+
(l_{\frac{n}{2}+1}-l_{\frac{n}{2}})& n\text{
even}\end{array}\right.
\end{align}

\noindent So, squaring this, one obtains a sum over two different
types of terms, in particular
\begin{align}
(l_a-l_b)^2\qquad\text{and}\qquad(l_a-l_b)(l_c-l_d),
\end{align}

\noindent where in the first term $a$ and $b$ are different and in
the second term all $a$, $b$, $c$ $d$ are different. Just because
of this, terms of the last kind cancel when summing over $\Sym_n$:
\begin{align}
\sum_{\s\in\Sym_n}(l_{\s(a)}-l_{\s(b)})(l_{\s(c)}-l_{\s(d)})\;=\;0.
\end{align}

\noindent So we can only keep the terms of the first kind and,
with (\ref{Gl:LustigeZahlenreihe}), we obtain:
\begin{align}\nonumber
\sum_{\s\in\Sym_n}\left(\sum_{k=1}^n\Delta_k(\s(\vec
l))\right)^2&=\;\m_0^2\sum_{\s\in\Sym_n}\left((n-1)^2(l_{\s(n)}
-l_{\s(1)})^2+(n-3)^2(l_{\s(n-1)}-l_{\s(2)})^2+\ldots\right)\\[5pt]\label{Gl:SummationDieK_nGibt}
&\;=\;2k_n\m_0^2(n-2)!\;\sum_{k=1}^n\sum_{m=1}^n(l_k-l_m)^2,
\end{align}

\noindent where
\begin{align}
k_n\;=\;\left\{\begin{array}{ll}
1^2\,+\,3^2+\,\cdots\,+\,(n-1)^2 & n\text{ even}\\[5pt]2^2\,+\,4^2\,+\,\cdots\,+\,(n-1)^2 & n\text{ odd}\end{array}\right..
\end{align}

\noindent A short analysis reveals that
$k_n=\frac{1}{6}(n-1)n(n+1)$, independently of whether $n$ is even
or odd. So, combining the results from
(\ref{Gl:ExpansionCarriedOut}) and (\ref{Gl:SummationDieK_nGibt}),
we get
\begin{align}
\nbra\,\hat
M^{2n}\,\mket\;&\approx\;\frac{(-1)^{\NN}}{256^n}\left(\begin{array}{c}8n\\4n+\NN\end{array}\right)\left(
\frac{\n+\m}{2}\right)^{2n}\\[5pt]\nonumber
\;&\quad+\;2\m_0^2\frac{(-1)^{\NN}}{256^n}\frac{n+1}{3}\left(\frac{\n+\m}{2}\right)^{2n-2}\sum_{\scriptsize
\begin{array}{c}l_1,\ldots,l_n\in\Z\\\sum_k
l_k=\frac{\n-\m}{4\m_0}\end{array}}\left[\prod_{k=1}^n\left(\begin{array}{c}8\\4+l_k\end{array}\right)\right]
\sum_{k,m}(l_k-l_m)^2\\[5pt]\nonumber
&\quad-\;\frac{(-1)^{\NN}}{256^nn!}\left(\frac{\n+\m}{2}\right)^{2n-2}\sum_{\scriptsize
\begin{array}{c}l_1,\ldots,l_n\in\Z\\\sum_k
l_k=\frac{\n-\m}{4\m_0}\end{array}}\left[\prod_{k=1}^n\left(\begin{array}{c}8\\4+l_k\end{array}\right)\right]
\sum_{\s\in\Sym_n}\sum_{k=1}^n\Delta_k(\s(\vec l))^2.
\end{align}

\noindent Now
\begin{align}
\sum_{k,m}(l_k-l_m)^2\;=\;n\sum_{k=1}^nl_k^2\;-\;\left(\sum_{k=1}^nl_k\right)^2
\end{align}

\noindent and thus
\begin{align}\label{Gl:WichtigerSchritt}
\nbra\,\hat
M^{2n}\,\mket\;&\approx\;\frac{(-1)^{\NN}}{256^n}\left(\begin{array}{c}8n\\4n+\NN\end{array}\right)
\left(\frac{\n+\m}{2}\right)^{2n}\\[5pt]\nonumber
\;&\quad-\;2\frac{(-1)^{\NN}}{256^n}\frac{n+1}{3}\left(\begin{array}{c}8n\\4n+\NN\end{array}\right)\left(\frac{\n+\m}
{2}\right)^{2n-2}\left(\frac{\n-\m}{4}\right)^{2}\\[5pt]\nonumber
\;&\quad+\;2\m_0^2\frac{(-1)^{\NN}}{256^n}\frac{n(n+1)}{3}\left(\frac{\n+\m}{2}\right)^{2n-2}
\sum_{\scriptsize\begin{array}{c}l_1,\ldots,l_n\in\Z\\\sum_k
l_k=\frac{\n-\m}{4\m_0}\end{array}}\left[\prod_{k=1}^n\left(\begin{array}{c}8\\4+l_k\end{array}\right)
\right]\sum_{k=1}^nl_k^2\\[5pt]\nonumber
&\quad-\;\frac{(-1)^{\NN}}{256^nn!}\left(\frac{\n+\m}{2}\right)^{2n-2}\sum_{\scriptsize
\begin{array}{c}l_1,\ldots,l_n\in\Z\\\sum_k
l_k=\frac{\n-\m}{4\m_0}\end{array}}\left[\prod_{k=1}^n\left(\begin{array}{c}8\\4+l_k\end{array}\right)\right]
\sum_{\s\in\Sym_n}\sum_{k=1}^n\Delta_k(\s(\vec l))^2
\end{align}

\noindent Let us now evaluate the last line in this expression.
With a bit of algebra one can show (for $n>1$):
\begin{align}
\frac{1}{\m_0^2n!}\sum_{\s\in\Sym_n}\sum_{k=1}^n\,\Delta_k(\s(\vec
l))^2\;=\;(n-1)\left(\sum_{k=1}^nl_k^2\right)\,+\,\frac{2}{n!}\frac{(n+1)(n-1)^2}{6}\left(\sum_{k<
j}l_kl_j\right),
\end{align}

\noindent of which we only keep the first term, because the second
one is small compared to it for large $n$.

\begin{align}\label{Gl:WichtigerSchritt2}
\nbra\,\hat
M^{2n}\,\mket\;&\approx\;\frac{(-1)^{\NN}}{256^n}\left(\begin{array}{c}8n\\4n+\NN\end{array}\right)
\left(\frac{\n+\m}{2}\right)^{2n}\\[5pt]\nonumber
\;&\quad-\;2\frac{(-1)^{\NN}}{256^n}\frac{n+1}{3}\left(\begin{array}{c}8n\\4n+\NN\end{array}\right)\left(\frac{\n+\m}
{2}\right)^{2n-2}\left(\frac{\n-\m}{4}\right)^{2}\\[5pt]\nonumber
\;&\quad+\;\m_0^2\frac{(-1)^{\NN}}{256^n}\;\frac{2n^2-n+3}{3}\;\left(\frac{\n+\m}{2}\right)^{2n-2}
\sum_{\scriptsize\begin{array}{c}l_1,\ldots,l_n\in\Z\\\sum_k
l_k=\frac{\n-\m}{4\m_0}\end{array}}\left[\prod_{k=1}^n\left(\begin{array}{c}8\\4+l_k\end{array}\right)
\right]\sum_{k=1}^nl_k^2.
\end{align}

\noindent The constrained sum can be evaluated by means of the
same trick that has been applied to calculate the constrained sum
(\ref{Gl:AusgangsGleichungMitSumme}). Without showing all the
steps again, one obtains:
\begin{align}\nonumber
\sum_{\scriptsize\begin{array}{c}l_1,\ldots,l_n\in\Z\\\sum_k
l_k=\frac{\n-\m}{4\m_0}\end{array}}&\left[
\prod_{k=1}^n\left(\begin{array}{c}8\\4+l_k\end{array}\right)\right]\sum_{k=1}^nl_k^2\\[5pt]\label{Gl:SummeUeberQuadrate}
&\;=\;n\sum_{\scriptsize\begin{array}{c}l_1,\ldots,l_n\in\Z\\\sum_k
l_k=\frac{\n-\m}{4\m_0}\end{array}}\left[
\prod_{k=1}^n\left(\begin{array}{c}8\\4+l_k\end{array}\right)\right]l_1^2\\[5pt]\nonumber
&\;=\;n\frac{d^2}{da^2}_{\big|a=0}\int_0^{2\pi}\frac{db}{2\pi}\,e^{-4a}\left(1+e^{a+ib}\right)^8\left(1+e^{ib}
\right)^{8(n-1)}e^{-ib\NN}\\[5pt]\nonumber
&\;=\;n\left(
16\left(\begin{array}{c}8n\\4n+\NN\end{array}\right)\;-\;56\left(\begin{array}{c}8n-1\\4n-1+\NN\end{array}\right)\;
+\;56\left(\begin{array}{c}8n-2\\4n-2+\NN\end{array}\right)
\right).
\end{align}

\noindent With formula (\ref{Gl:RelateBinomialCoefficients}) one
can calculate
\begin{align}\label{Gl:KleinesPolynom}
16\left(\begin{array}{c}8n\\4n+\NN\end{array}\right)\;-&\;56\left(\begin{array}{c}8n-1\\4n-1+\NN\end{array}\right)\;
+\;56\left(\begin{array}{c}8n-2\\4n-2+\NN\end{array}\right)\\[5pt]\nonumber
&\;=\;\left(16-\frac{112n}{8n-1}+\frac{7}{(8n-1)n}\left(\NN\right)^2\right)\left(\begin{array}{c}8n\\4n+\NN\end{array}
\right).
\end{align}

\noindent Inserting (\ref{Gl:SummeUeberQuadrate}) and
(\ref{Gl:KleinesPolynom}) into (\ref{Gl:WichtigerSchritt}) gives:
\begin{align}\label{Gl:FinaleApproximationFormel}
\nbra\,\hat
M^{2n}\,\mket\;&\approx\;\frac{(-1)^{\NN}}{256^n}\left(\begin{array}{c}8n\\4n+\NN\end{array}\right)
\left(\frac{\n+\m}{2}\right)^{2n}\\[5pt]\nonumber
\;&\quad+\;\frac{(-1)^{\NN}}{256^n}\left(\begin{array}{c}8n\\4n+\NN\end{array}\right)\;
\m_0^2\left(
\frac{\n+\m}{2}\right)^{2n-2}\left(\frac{n(n-1)}{8n-1}\,\frac{2n^2-n+3}{3} \right)\\[5pt]\nonumber
\;&\quad+\;\frac{(-1)^{\NN}}{256^n}\left(\begin{array}{c}8n\\4n+\NN\end{array}\right)\m_0^2\left(\frac{\n+\m}{2}
\right)^{2n-2}\left(\NN\right)^2\;\frac{1}{3}\,\frac{n-1}{8n-1}(16n^2+16n+21).
\end{align}

\noindent In our calculations we have assumed $n>1$, but formula
(\ref{Gl:FinaleApproximationFormel}) also produces the right
result for $n=1$, as one can readily see. Note that the binomial
coefficient for $n=0$ produces a $0$ for $\n\neq \m$. For $\n=\m$
and $n=0$ the binomial coefficient gives a $1$, as does the whole
expression (\ref{Gl:FinaleApproximationFormel}). So for $n=0$ the
expression (\ref{Gl:FinaleApproximationFormel}) equals
$\d_{\n\m}$, as expected. Hence, the approximation formula is
valid
even for all $n\geq 0$.\\


\subsection{The physical inner product}

\noindent We see that, considering $\n-\m$ being small compared to
$\n+\m$, the first term in (\ref{Gl:FinaleApproximationFormel}) is
large compared to the second and the third, although not uniformly
for all $n$. Since, to compute the rigging map, one has to sum
over all the orders of $n$, the contribution of the two correction
terms could be considerable, if $\n+\m$ is too small. So we see
that for small $\n+\m$, the major contribution might not come from
the zeroth order term, but from the corrections. Fortunately, one
can calculate the corrections to the rigging map order by order,
which we will do now. The zeroth order term is
\begin{align}
\nbra\,\hat
M^{2n}\,\mket\;&\approx\;\frac{(-1)^{\NN}}{256^n}\left(
\begin{array}{c}8n\\4n+\NN\end{array}\right)\left(\frac{\n+\m}{2}\right)^{2n}.
\end{align}

\noindent The rigging map is given  by :
\begin{align}\label{Gl:RiggingMap}
\eta[\nket]\,\mket&\;=\;\lim_{t\to
0}\;\frac{\sum_{n=0}^{\infty}\frac{(-1)^n}{n!t^{n}} \nbra\,\hat
M^{2n}\,\mket } {\sum_{n=0}^{\infty}\frac{(-1)^n}{n!t^{n}} \langle
\n_0|\,\hat M^{2n}\,|\n_0\rangle},
\end{align}

\noindent where $|\n_0\rangle$ is a reference vector. With the
abbreviations
\begin{align}\label{Gl:Abbreviations}
a\;=\;\left|\frac{\n+\m}{32\m_0\sqrt t}\right|\;,\qquad N=\NN.
\end{align}

\noindent we can make use of the integral
(\ref{Gl:ChristiansFormelohneSinus})
\begin{align}
\sum_{n=0}^{\infty}\frac{(-1)^n}{2n!}\,a^{2n}\,&\frac{\G(8n+1)}{\G(4n+1+N)\G(4n+1-N)}\\[5pt]\nonumber
&\;=\;\frac{1}{2\pi i}\int_{-i\infty-\e}^{i\infty-\e}
ds\;\G(-s)\,a^{2s}\,\frac{\G(8s+1)}{\G(4s+1+N)\G(4s+1-N)},
\end{align}

\noindent where the integration contour is chosen to pass slightly
left from the pole at $0$. We now apply a coordinate
transformation $8s+1=-r$, so we get
\begin{align}
\frac{1}{2\pi i}\int_{-i\infty-\e}^{i\infty-\e}&
ds\;\G(-s)\,a^{2s}\,\frac{\G(8s+1)}{\G(4s+1+N)\G(4s+1-N)}\\[5pt]\nonumber
&\;=\;-\frac{1}{16\pi
i}\int_{i\infty-1+\e}^{-i\infty-1+\e}dr\;\G(-r)\,a^{-\frac{r+1}{4}}\,\frac{\G\left(\frac{r+1}{8}\right)}{\G\left(\frac{1}{2}-\frac{r}{8}+N\right)\G\left(\frac{1}{2}-\frac{r}{8}-N\right)}.
\end{align}

\noindent Observing that the integrand has no poles in
$[-1+\e,-\e]\times i\R$, we can shift the integration curve and
obtain
\begin{align}
\frac{1}{16\pi
i}&\int_{-i\infty-\e}^{i\infty-\e}dr\;\G(-r)\,a^{-\frac{r+1}{4}}\,
\frac{\G\left(\frac{r+1}{8}\right)}{\G\left(\frac{1}{2}-\frac{r}{8}+N\right)\G\left(\frac{1}{2}-\frac{r}{8}-N
\right)}\\[5pt]\noindent
&\;=\;\frac{1}{8}\,a^{-\frac{1}{4}}\;\sum_{n=0}^{\infty}\frac{(-1)^n}{n!}\,a^{-\frac{n}{4}}\,
\frac{\G\left(\frac{n+1}{8}\right)}{\G\left(\frac{1}{2}-\frac{n}{8}+N\right)\G\left(\frac{1}{2}-\frac{n}{8}-N\right)}.
\end{align}

\noindent This means that, reinserting $a$ and $N$ from
(\ref{Gl:Abbreviations}), only the $n=0$-Term survives in the
limit:
\begin{align}\nonumber
\lim_{t\to
0}\;t^{-\frac{1}{8}}\;&\sum_{n=0}^{\infty}\frac{(-1)^n}{n!}\;\left(\frac{\n+\m}{32\m_0t}\right)^{2n}\;
\frac{\G(8n+1)}{\G(4n+1+\NN)\G(4n+1-\NN)}\\[5pt]
&\;=\;\frac{1}{8}\left|\frac{32\m_0}{\n+\m}\right|^{\frac{1}{4}}\;\frac{\G\left(\frac{1}{8}\right)}
{\G\left(\frac{1}{2}+\NN\right)\G\left(\frac{1}{2}-\NN\right)}.
\end{align}

\noindent Together with
\begin{align}
\frac{1}{\G\left(\frac{1}{2}+\NN\right)\G\left(\frac{1}{2}-\NN\right)}\;=\;\frac{1}{\pi}\;(-1)^{\NN}
\end{align}

\noindent and (\ref{Gl:RiggingMap}) we arrive at the result
\begin{align}\label{Gl:FinaleFormelOhneKorrektur}
\eta[\nket]\,\mket\;\approx\;\left|\frac{2\n_0}{\n+\m}\right|^{\frac{1}{4}}.
\end{align}

\noindent The above calculation can be carried through likewise
with the correction terms in (\ref{Gl:FinaleApproximationFormel}),
which yields, after some careful analysis:
\begin{align}\label{Gl:FinaleFormelMitKorrektur}
\eta[\nket]\,\mket\;&\approx\;\left|\frac{2\n_0}{\n+\m}\right|^{\frac{1}{4}}\\[5pt]\nonumber
&\;+\;\left|\frac{2\m_0}{\n+\m}\right|^{\frac{9}{4}}\;\left(\frac{\n_0}{\m_0}\right)^{\left(\frac{1}{4}\right)}\;\left(\frac{231}{64}+\frac{303}{256}\left(\NN\right)^2
\right)\\[5pt]\nonumber
&\;-\;\left|\frac{2\m_0}{\n+\m}\right|^{\frac{9}{4}}\;\left(\frac{\n_0}{\m_0}\right)^{\left(\frac{1}{4}\right)}\;\frac{\G\left(\frac{7}{8}\right)^2}{\pi
\csc\left(\frac{\pi}{8}\right)}\frac{1}{\frac{1}{4}-\left(\NN\right)^2}\left(\frac{217}{4}-\frac{217}{16}\left(\NN\right)^2
\right).
\end{align}

\noindent For the calculations with the corrections, the choice of
$\n_0$ results in a choice of normalization of the rigging map.
However, the dependence of the normalization on $\n_0$ is
complicated. In (\ref{Gl:FinaleFormelMitKorrektur}) a particular
normalization has been chosen, to make the formula comparable to
(\ref{Gl:FinaleFormelOhneKorrektur}).

\subsection{Discussion}

\noindent First note that the choice of the reference vector
$|\n_0\rangle$ is only a choice of normalization of our rigging
map. It does not even have to be from the same sector
$\d\in[0,4\m_0)$ than $\n$ and $\m$. Note further that in our
approximation (\ref{Gl:EntwicklungInDeltaK}) we assumed that
$\n+\m$ is large compared to $\n-\m$. In particular, the zeroth
order approximation (\ref{Gl:FinaleFormelOhneKorrektur}) cannot be
trusted if this condition is violated and, as one can see, it even
diverges if $\m$
approaches $-\n$. We will now discuss this formula for the sectors $\d=0$ and $\d\neq 0$ separately.\\

$\d\in(0,\,4\m_0)$:\\

 The rigging map we have produced suggests, as a zeroth
order approximation, that
\begin{align}\label{Gl:TheFinalApproximateSolution}
\eta[\nket]\;\approx\;\sum_{\l\in\n+4\m_0\Z}\frac{1}{|\l+\n|^{\frac{1}{4}}}\;\langle\l|,
\end{align}

\noindent where the result should only be trusted, where
$\l\approx \n$, and both $\l$ and $\n$ are large.

We know that by construction (\ref{Gl:RiggingMapEquivalence}),
$\eta[\nket]$ solves the master constraint, i.e.
\begin{align}
\eta[\nket]\,\circ\,\hat M\;=\;0.
\end{align}

\noindent In chapter \ref{Ch:KapitelEins} we calculated the
solutions to the constraint $\hat C$ explicitly. There we saw that
in each sector $\d\in(0,\,4\m_0)$ there are two solutions
(\ref{Gl:SolutionsDelta}), one of which is bounded and one of
which is unbounded. For all sectors $\d\neq 0$, the unbounded
solution goes like $|\l|^{\frac{1}{2}}$, whereas the bounded one
goes like $|\l|^{-\frac{1}{2}}$ for $|\l|\to\infty$. Furthermore,
the master constraint $\hat M$ has two further solutions in each
sector, given by (\ref{Gl:AdditionalSolutions}), which behave
asymptotically like $|\l|^{\frac{3}{2}}$ and $|\l|^{\frac{5}{2}}$,
respectively.

For fixed $\nket$, the approximation stays bounded in the regime
where it is supposed to be a good approximation to the real
$\eta[\nket]$, and goes asymptotically like $|\l|^{-\frac{1}{4}}$.
Although this is not quite the right asymptotical behaviour, it is
safe to assume that the true $\eta[\nket]$ is the \emph{bounded}
solution
\begin{align}\label{Gl:TheFinalSolution}
\eta[\nket]\;=\;\sum_{\l\in\d+4\m_0\Z}\frac{1}{|\l+\m_0|^{\frac{3}{2}}\,-\,|\l-\m_0|^{\frac{3}{2}}}\,\langle\l|,
\end{align}

\noindent i.e. the one going like $|\l|^{-\frac{1}{2}}$. So,
although the approximation (\ref{Gl:TheFinalApproximateSolution})
does not recover the exact asymptotical behaviour of the physical
solution (\ref{Gl:TheFinalSolution}), it is good enough to
identify the right solution (\ref{Gl:TheFinalSolution}) among the
four possible ones (\ref{Gl:SolutionsDelta}) and
(\ref{Gl:AdditionalSolutions}) in each sector $\d\in(0,\,4\m_0)$.

 Numerical comparison of the different approximations
(\ref{Gl:AusgangsGleichungMitSumme}) and
(\ref{Gl:FinaleApproximationFormel}) with the real values for
$\nbra\hat M^{2n}\mket$ for different $n$ indicates that the
approximation provided solely by the coherent states
(\ref{Gl:AusgangsGleichungMitSumme}) is quite good, whereas
(\ref{Gl:FinaleApproximationFormel}) is not quite that accurate.
This shows that the reason for the slightly different asymptotic
behaviour of the final approximation of the physical solution
compared to the one expected is caused by further simplifying the
formula in chapter \ref{Ch:FurtherApproximations}, in particular
by by going over from (\ref{Gl:AusgangsGleichungMitSumme2}) to
(\ref{Gl:StepWhereFurtherApproximationIsMade}). This was necessary
in order to bring the expressions into a form such that further
calculations were possible, but it happened at the cost of
accuracy. Still, the approximation formula can in principle be
improved by computing and adding further corrections, such as
happened with the first order corrections in
(\ref{Gl:FinaleFormelMitKorrektur}).

 Note that in particular, although the rigging map was
defined by the direct integral decomposition of $\hat M$
(\ref{Gl:RiggingMapEquivalence}), the physical Hilbert space
contains only solutions of $\hat M$ that are also solutions of
$\hat C$. This shows that, although one had replaced $\hat C$ by
$\hat M$, no information is lost, and hence the master constraint
programme proved to be the right tool to define the physical
Hilbert space in the presence of a non-self-adjoint constraint.\\

Note further that since, analogously to the case of chapter
\ref{Ch:RiemannianCosmology}, in each sector $\d\in(0,\,4\m_0)$
there is exactly one solution of nonzero physical norm (hence all
other solutions in the same sector are spurious), the Hilbert
space is again non-separable, in particular it contains
\begin{align}
\bigoplus_{\d\in(0,\,4\m_0)}\C\;\subset\;\H_{phys}.
\end{align}

\noindent Hence, in the sectors $\d\in(0,\,4\m_0)$, the result for
the physical Hilbert space provided by the master constraint is
the same as for the simplified self-adjoint constraint
(\ref{Gl:SimpleConstraint}), which in fact is classically
equivalent to (\ref{Gl:NotSoSimpleConstraint}) apart from $p=0$.
So the more complicated, non-self-adjoint constraint
(\ref{Gl:NotSoSimpleConstraint}), together with the master
constraint leads to the same physical Hilbert space as the
simplified constraint (\ref{Gl:SimpleConstraint}).\\

$\d=0$:\\

The case of $\d=0$ has to be treated differently than the case
$\d\in(0,\,4\m_0)$. The reason for this is that the vector
$|0\rangle$ is contained in $\H_{\d=0}$, which corresponds by
(\ref{Gl:EigenvectorOfP}) to the classical singularity $p=0$,
 defining the points in phase space where the two constraints
(\ref{Gl:SimpleConstraint}) and (\ref{Gl:NotSoSimpleConstraint})
differ classically. Indeed, $|0\rangle$ is an eigenvector with
eigenvalue 0 of (\ref{Gl:NotSoSimpleConstraint})
\begin{align}
\hat C\,|0\rangle\;=\;\left(\frac{\sin \m_0\hat
c}{\m_0}\right)^2\;\sgn\,\hat
p\,\widehat{\sqrt{|p|}}\,|0\rangle\;=\;0,
\end{align}

\noindent but not of (\ref{Gl:SimpleConstraint}). Hence the
solutions of (\ref{Gl:NotSoSimpleConstraint}), given by
(\ref{Gl:SolutionsDeltaZero}), have a different structure than
(\ref{Gl:SolutionsDelta}): There is no bounded solution with
asymptotical behaviour $|\l|^{-\frac{1}{2}}$, instead there is the
eigenvector $|0\rangle$. This eigenvector constitutes $\H_0^{pp}$
and has to be excluded from the following discussion, which deals
with $\H^{ac}_0$ only (see chapter \ref{Ch:PHYS}).

Although there is no analogue of (\ref{Gl:TheFinalSolution}) in
the sector with $\d=0$, the approximation that has been calculated
gives
\begin{align*}
\eta[\nket]\;\approx\;\sum_{\l\in\n+4\m_0\Z}\frac{1}{|\l+\n|^{\frac{1}{4}}}\;\langle\l|,
\end{align*}

\noindent even for $\n\in 4\m_0\Z$, i.e. in the sector $\d=0$. So
the approximation scheme treats all $\d\in[0,\,4\m_0)$ equally. In
the case for $\d\neq 0$, (\ref{Gl:TheFinalApproximateSolution})
has been interpreted as an approximation to the solution
(\ref{Gl:TheFinalSolution}), which, as we have seen, does not
exist in this sector. Thus, it is not clear which solution the
rigging map approximates, if any. We attempt to interpret this as
follows:

The disappearance of an analogue to the solution
(\ref{Gl:TheFinalSolution}) for $\d=0$ can be traced back to the
appearance of $\sgn\,\hat p \widehat{\sqrt{|p|}}$ in the
constraint (\ref{Gl:NotSoSimpleConstraint}), which results in an
eigenvector of $\hat C$ (hence $\hat M$) with eigenvalue 0. The
eigenvector $|0\rangle$ of $\hat C$ has to be removed from the
analysis, since discrete and continuous spectrum have to be
treated separately in the direct integral
decomposition\footnote{In particular, one has to perform the
direct integral decomposition separately for each part of the
spectrum, leading to a different choice of normalization vector
$|\n_0\rangle$ in (\ref{Gl:RiggingMapEquivalence})}. Hence, the
Hilbert space $\H_0$ has to be replaced by ${\rm
span}\,\{\nket\;|\;\n\in4\m_0\Z\backslash\{0\}\}$, which is a
sub-Hilbert space in $\H_0$ of codimension 1. The coherent states,
however, are defined on all of $\H_0$, in particular the
projection to $|0\rangle$ is nonzero for all coherent states in
$\H_0$, i.e. states (\ref{Gl:CoherentStates}) with $\d=0$. So, by
using the coherent states for the semiclassical approximation on
the sector $\d=0$, one is extending the space one is allowed to
look for solutions. So we have to conclude that the approximation
provided by the coherent states are not appropriate for the sector
$\d=0$, since, strictly speaking, there \emph{are} no coherent
states living on this sector. Thus, the approximation cannot be
trusted at all, and we conclude that there is \emph{no} physical
solution in the sector $\d=0$. Hence, the whole physical Hilbert
space is
\begin{align}
\H_{phys}\,=\;\H_0^{pp}\oplus\H^{ac}_{0}\;=\;{\rm span
}\{|0\rangle\}\,\oplus\,\bigoplus_{\d\in(0,\,4\m_0)}
\C.\;=\;\bigoplus_{\d\in[0,\,4\m_0)} \C
\end{align}

\section{Summary and conclusion}

\noindent In this article, we computed the physical inner product
of a simple cosmological model. We did so in two different ways:
Firstly, we rewrote the classical side before quantization, in
order to arrive at a much simpler constraint on the quantum side,
the physical inner product of which could be computed easily. In
each Hilbert space $\H_{\d}$, there are a bounded and an unbounded
solution to the constraint, and the bounded one was constituting
the physical Hilbert space $\H_{phys}$, while the other solution
was "spurious", i.e. not contained in it. In the model calculated
$H_{phys}$ is non-separable. This unphysical feature is a result
of the quantization procedure, which brakes scale-invariance of
the classical theory. Still, this was no particular problem for
our calculations.

 Secondly, we attempted
to compute the physical inner product directly, without rewriting
of the classical side. This resulted in a system which is
classically equivalent up to points with $p=0$ on the phase space.
The constraint of the new system is non-symmetric, which is why we
employed the master constraint method to arrive at a self-adjoint
constraint. Since the formulae for this operator were too
complicated to compute exactly, we relied on the complexifier
coherent states to approximate the physical inner product. This
has led to interesting results:

Similarly to the case already observed, on each sector
$\d\in(0,4\mu_0)$ there are a bounded and an unbounded solution,
showing that rewriting on the classical side did not change the
quantum theory significantly on the parts of the phase space where
the two classical theories are equivalent. The sector $\d=0$
contained the eigenvector of $\hat p$ belonging to vanishing
momentum $p=0$, which defines the phase-space points where the two
classical theories do not agree, and indeed the solution structure
of the quantum constraints changed significantly. What did not
change is the physical Hilbert space. Although now containing an
eigenvector, $|0\rangle$, the resulting physical Hilbert space
$\H_{phys}$ was isomorphic to the one computed from the simpler
model. So rewriting the classical theory changed the structure of
the solution space in quantum theory at the particular points,
where the classical theories disagree, but, after going over to
the physical Hilbert space, leading to the right result.

The approximation of the physical inner product with the help of
the complexifier coherent states was good enough to indicate that
for each sector $\d\in(0,4\m_0)$ the bounded solution was chosen
to constitute the physical inner product, while all the unbounded
solutions turned out to be spurious. This demonstrated firstly
that going over to the master constraint did not change the
quantum behaviour. In particular, although there are many more
solutions to the master constraint $\hat M$ than to the original
constraint $\hat C$, computing the physical inner product with
$\hat M$ still resulted in a physical Hilbert space consisting of
solutions to $\hat C$ only. Hence the additional solutions
provided by the master constraint did not contribute to the
physical Hilbert space, showing the strength of the master
constraint programme.

This demonstrates that the complexifier coherent states provide a
way to approximate physical inner products. Furthermore, the
corrections to the approximated rigging map can in principle be
calculated order by order (the "perturbation parameter" in our
case being the distance to the region of phase space where the
approximation was to hold best) and be added to the previously
obtained result, granting a way to improve the approximation by
adding more and more correction terms.

\section{Acknowledgements}\nonumber

\noindent BB would like to thank Christian Hillmann, for pointing
out the significance of formula
(\ref{Gl:ChristiansFormelohneSinus}).

\end{document}